\documentclass[
 reprint,
 superscriptaddress,
 nofootinbib,
 amsmath,amssymb,
 aps,
 prl,
 longbibliography,
]{revtex4-2}

\usepackage{graphicx}
\usepackage{dcolumn}
\usepackage{bm}
\usepackage{natbib}
\usepackage{hyperref}
\hypersetup{colorlinks=true,linkcolor=blue,citecolor=blue,filecolor=blue,urlcolor=blue}
\usepackage{indentfirst}

\begin{document}

\title{Current-induced sliding motion in a helimagnet MnAu$_2$}

\author{Yuta~Kimoto}
\email{yuta.kimoto.t7@dc.tohoku.ac.jp}
\author{Hidetoshi~Masuda}
\affiliation{Institute for Materials Research, Tohoku University, Sendai 980-8577, Japan}
\author{Takeshi~Seki}
\affiliation{Institute for Materials Research, Tohoku University, Sendai 980-8577, Japan}
\affiliation{Center for Science and Innovation in Spintronics (CSIS), Core Research Cluster, Tohoku University, Sendai 980-8577, Japan}
\author{Yoichi~Nii}
\affiliation{Institute for Materials Research, Tohoku University, Sendai 980-8577, Japan}
\author{Jun-ichiro~Ohe}
\affiliation{Department of Physics, Toho University, Funabashi 274-8510, Japan}
\author{Yusuke~Nambu}
\affiliation{Institute for Materials Research, Tohoku University, Sendai 980-8577, Japan}
\author{Yoshinori~Onose}
\affiliation{Institute for Materials Research, Tohoku University, Sendai 980-8577, Japan}

\begin{abstract}
We found signatures of current-induced sliding motion in helimagnetic $\mathrm{Mn}\mathrm{Au}_2$ thin films. An abrupt change in differential resistivity occurred at a threshold bias current in the helimagnetic state, whereas it was absent in the induced ferromagnetic state. Broadband voltage noise also emerged above the threshold current in the helimagnetic state. Based on the similarity to canonical charge/spin density wave systems, we ascribed the origin of these phenomena to the sliding motion of the helimagnetic structure.
\end{abstract}

\maketitle
In charge- or spin-ordered states, a large electric field sometimes drives the sliding motion of the ordered periodic structure. For example, in quasi-one-dimensional systems, a charge/spin density wave (CDW/SDW) exhibits sliding motion under a large electric field \cite{gruner_CDW_transport,gruner_dynamics_SDW,monceau_electronic}. Theoretically, the sliding motion is described as the zero-energy excitation in the gapless Goldstone (phason) mode \cite{lee_conductivity_Sliding}, but in reality it is pinned by impurities and commensurability, and therefore a certain magnitude of electric field is needed to drive the sliding motion. The differential resistivity, which is alternating current resistivity under a DC electric field, is abruptly decreased at the threshold DC electric field. The decrease is ascribed to the onset of sliding motion \cite{gruner_CDW_transport,gruner_dynamics_SDW}. Similar sliding phenomena have been observed in Wigner crystals on a liquid $^4$He surface \cite{shirahama_dynamical_LHe4} and strongly-correlated charge- and spin-ordered states in transition metal oxides \cite{yamanouchi_dielectric_sliding,cox_manganite_sliding,blumberg_SCOsliding}.

In non-collinear magnetic structures, an electric current effectively drives the sliding motion because the transverse component of spin moment carried from the adjacent magnetic site induces a magnetic torque, which is denoted as spin-transfer torque (STT) \cite{brataas_current-induced_torque}. For example, a ferromagnetic domain wall is driven by an electric current larger than ${10}^{11}$ A/m$^2$ in the metallic case  \cite{yamaguchi_FMDW_Observation,koyama_intrinsic_pinning} or ${10}^{8}$ A/m$^2$ in the semiconducting case \cite{yamanouchi_current-induced_GaMnAsDW}. The topological non-collinear spin texture known as Skyrmion also shows sliding motion by STT with much less threshold electric current \cite{jonietz_STT_MnSi,schulz_emergent_skyrmion,yokouchi_current-induced_skyrmion,sato_SkyrmionNoise,he_antiSKy_sliding,birch_Sky_sliding}.

\begin{figure}[t]
\includegraphics[width=\linewidth]{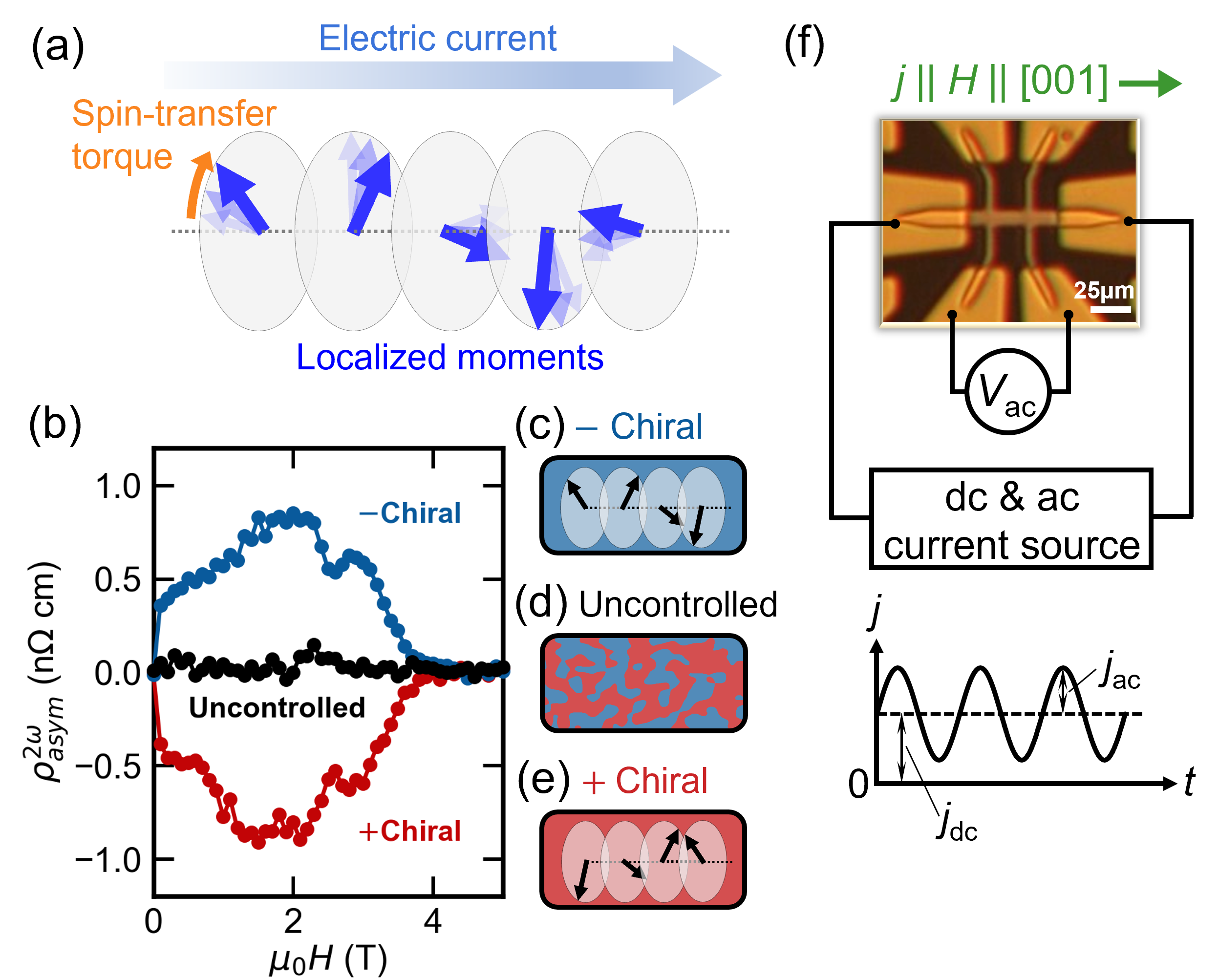}
\caption{(a) Schematic image of sliding motion of helimagnetic structure. (b) Nonreciprocal resistivity $\rho_\mathrm{asym}^{2\omega}$ for $\pm$chiral and chirality-uncontrolled states at 250 K. (c)-(e) Schematic images of $\pm$chiral and chirality-uncontrolled states. (f) The experimental setup for the differential resistivity measurement.}
\label{fig:FIG1}
\end{figure}

The helical spin structure is one type of the non-collinear magnetic structure. Many theoretical papers have predicted the sliding motion of the helical spin structure \cite{tserkovnyak_adiabatic_pumping,wessely_current_dynamics_helimag,wessely_STT_helical,iwasaki_universal_HLSk,del_archimedean,ustinov_current-induced_rotation_helimag,kurebayashi_pumping,xie_sliding_SkXHL}. A schematic image of the sliding motion of a helimagnetic structure is depicted in \autoref{fig:FIG1}(a). When the electric current $j$ is applied along the helical axis in a helical spin structure, the adiabatic STT expressed as $\bm{\tau}_{\mathrm{STT}}\propto(\bm{j}\cdot\bm{\nabla})\bm{S}$, in which $\bm{S}$ is the spatially dependent spin moment, tends to rotate the helical spin structure around the helical axis \cite{wessely_current_dynamics_helimag}. The collective rotation of the helimagnetic structure corresponds to the sliding motion. While the picture of sliding motion in the helimagnetic structure is clear, experimental observations have not been reported yet. In this Letter, we report experimental signatures of sliding motion in a helimagnet $\mathrm{Mn}\mathrm{Au}_2$.

$\mathrm{Mn}\mathrm{Au}_2$ crystallizes in a centrosymmetric tetragonal crystal structure with the space 
group $I4/mmm$ \cite{hall_MnAu2structure}. The Mn magnetic moments exhibit a helical magnetic order with a 
helical plane perpendicular to the propagation vector, which is $q=(0, 0, 0.28)$ in the reciprocal 
lattice unit, corresponding to a helical pitch of 3.1 nm \cite{herpin_MnAu2neutron}. The transition 
temperature is reported to be $T_c\approx360$ K for a bulk crystal \cite{meyer_MnAu2Tc}  and $T_c\approx340$ K for a thin film \cite{masuda_chirality_control_MnAu2}. To observe the sliding motion of the helical spin structure, we used single-crystalline helimagnetic $\mathrm{Mn}\mathrm{Au}_2$ thin films. $\mathrm{Mn}\mathrm{Au}_2$ (100 nm) films were epitaxially grown on $\mathrm{Sc}\mathrm{Mg}\mathrm{Al}\mathrm{O}_4$ (10-10) substrates by sputtering. The $\mathrm{Mn}\mathrm{Au}_2$ thin films are stacked along the [110] direction so that the helical propagation vectors are parallel to the thin films. A capping layer of Ta (2 nm) was deposited on top of the $\mathrm{Mn}\mathrm{Au}_2$ films. Hall bar devices with channel length 25 $\mathrm{\mu}$m and width 10 $\mathrm{\mu}$m [\autoref{fig:FIG1}(f)] were fabricated using photolithography and Ar ion milling. The direction of the electric current was parallel to the $\mathrm{Mn}\mathrm{Au}_2$ [001] direction, which is along the helical propagation vector. Details of the sample fabrication are described elsewhere~\cite{masuda_chirality_control_MnAu2}. 

Previously, Masuda \textit{et al.} successfully controlled the chirality of a helical spin structure in $\mathrm{Mn}\mathrm{Au}_2$ thin films by the simultaneous application of an electric current and a magnetic field \cite{masuda_chirality_control_MnAu2}. Similarly to their work, we confirmed that chirality control works for the present sample, as shown in \autoref{fig:FIG1}(b). We swept the magnetic field along the [001] direction from $\pm$5 T to 0 T with the application of an electric current ($j_0=8.0\times10^9$ A/m$^2$) and then measured the nonreciprocal electronic transport that reflects the chirality. The sign of the nonreciprocal resistivity $\rho_\mathrm{asym}^{2\omega}$, which is the field-asymmetric component of second harmonic resistivity, depends on whether the magnetic field and electric current applied prior to the measurement were parallel or antiparallel. This indicates that the chirality was certainly controlled. Here, we define the chiral state controlled by the parallel electric current and magnetic field ($\mu_0H=+5$ T) with negative $\rho_\mathrm{asym}^{2\omega}$ as the $+$chiral state and that controlled by the antiparallel magnetic field ($\mu_0H=-5$ T) and current with positive $\rho_\mathrm{asym}^{2\omega}$ as the $-$chiral state. When the electric current had not been applied, the nonreciprocal transport was negligible. This indicates that the magnetic domains with positive and negative chiralities coexisted, as shown in \autoref{fig:FIG1}(d). This state is defined as the chirality-uncontrolled state. Subsequent to the chiral domain control, we measured differential resistivity to observe the signature of the sliding motion. A DC and AC current source (Keithley, 6221) was connected to the Hall-bar-shaped sample, as shown in \autoref{fig:FIG1}(f). The superposition of AC current $j_\mathrm{ac}$ and DC bias current $j_\mathrm{dc}$ were applied, and the induced AC electric field $E_\mathrm{ac}$ was measured. Then, we obtained the differential resistivity $\rho_{xx}=\frac{E_\mathrm{ac}}{j_\mathrm{ac}}$ as a function of $j_\mathrm{dc}$. The amplitude and the frequency of $j_\mathrm{ac}$ were $5.05\times10^8$ A/m$^2$ and 79.19 Hz, respectively. In the following, we discuss $\rho_{xx}$ for the three chiral domain states: the $+$chiral state [\autoref{fig:FIG1}(e)], the $-$chiral state [\autoref{fig:FIG1}(c)] and the chirality-uncontrolled state [\autoref{fig:FIG1}(d)].

\begin{figure}[t]
\includegraphics{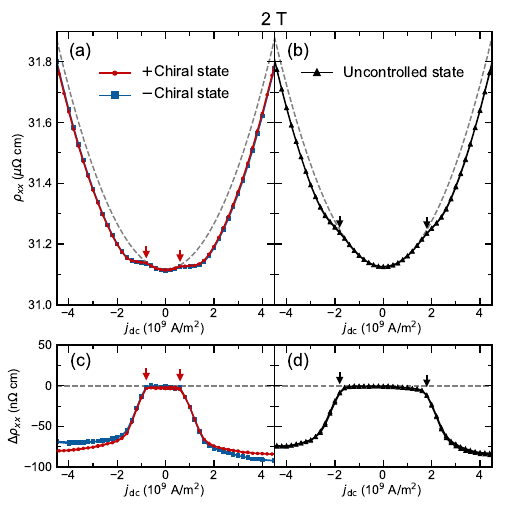}
\caption{(a, b) The differential resistivity in (a) chirality-controlled states with positive (red) and negative (blue) signs and (b) chirality-uncontrolled state at 250 K and 2 T. (c,~d) The differential resistivity subtracted from the quadratic background ($\Delta\rho_{xx}$) in (c) the chirality-controlled states and (d) that in the uncontrolled state at 250 K and 2 T.}
\label{fig:FIG2}
\end{figure}

\hyperref[fig:FIG2]{Figure~\ref*{fig:FIG2}} shows the real part of $\rho_{xx}$ as a function of $j_\mathrm{dc}$ for the three chiral domain states in the helimagnetic state ($T=$ 250 K, $\mu_0H=$ 2.0 T). Precisely speaking, in a finite magnetic field perpendicular to the helical plane, the magnetic moments are canted to the field direction, forming a conical magnetic structure. Because the imaginary part was always smaller than 0.5\% of the real part \cite{supp}, we only show the real part and ignore the imaginary part. In the chirality-uncontrolled state, $\rho_{xx}$ shows quadratic $j_\mathrm{dc}$ dependence in the low $j_\mathrm{dc}$ region, which is attributed to the Joule heating. The increase in temperature is approximately 2 K at $4\times10^9$ A/m$^2$. However, $\rho_{xx}$ shows kinks around $j_\mathrm{dc}=\pm2.0\times{10}^9$ A/m$^2$. In the high $j_\mathrm{dc}$ region, $\rho_{xx}$ shows quadratic $j_\mathrm{dc}$ dependence again. The heating effect should be monotonic as a function of bias current and cannot be the origin of kinks. To probe the kink structure more clearly, we subtract the quadratic $j_\mathrm{dc}$ dependent background as follows:
\begin{equation}
    \Delta\rho_{xx}(j_\mathrm{dc})=\rho_{xx}(j_\mathrm{dc})-\rho_{xx}(0 \mathrm{mA})\Big(1+\alpha j_\mathrm{dc}^2\Big),
\end{equation}
where $\alpha=1.2\times{10}^{-15}$ A$^{-2}$ m$^4$.
The same formula is used for the other magnetic field data, and coefficient $\alpha$ is determined so as to minimize the current dependence of $\Delta\rho_{xx}$ in the high-current region for all the field data. As shown in \autoref{fig:FIG2}(d), $\Delta\rho_{xx}$ clearly shows abrupt decreases at threshold currents $j_\mathrm{th}\approx\pm2.0\times10^9$ A/m$^2$.
In the chirality-controlled ($\pm$chiral) states, $\rho_{xx}$ shows kinks and $\Delta\rho_{xx}$ shows abrupt decreases similarly, but the threshold current is lower ($j_\mathrm{th}\approx\pm1.0\times10^9$ A/m$^2$), as shown in \hyperref[fig:FIG2]{Figs.~\ref*{fig:FIG2}}(a) and 2(c). These behaviors are independent of the sign of the chirality. As mentioned above, similar abrupt decreases of differential resistivity were reported in the CDW/SDW systems, which is ascribed to the sliding motion \cite{gruner_CDW_transport,gruner_dynamics_SDW}.

\begin{figure}[t]
\includegraphics{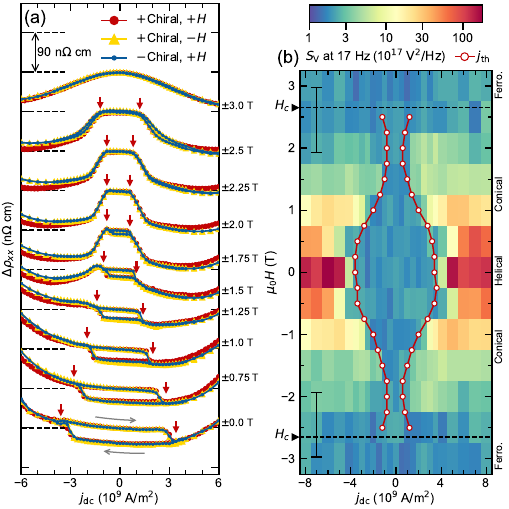}
\caption{(a) $\Delta\rho_{xx}$ at various positive magnetic fields $H$ for the $+$chiral (red circle) and -chiral (blue dot) states and that at various negative magnetic fields for $+$chiral state (yellow triangle) at 250 K. The red arrows denote threshold currents $j_\mathrm{th}$ at each magnetic field. (b) The magnetic field dependence of $j_\mathrm{th}$ at 250 K. Contour mapping of voltage noise at 17 Hz as a function of $j_\mathrm{dc}$ and magnetic field at 250 K is superimposed.}
\label{fig:FIG3}
\end{figure}

To further delve into the anomaly of $\Delta\rho_{xx}$, let us discuss the magnetic field dependence. \hyperref[fig:FIG3]{Figure~\ref*{fig:FIG3}}(a) shows the $j_\mathrm{dc}$ dependence of $\Delta\rho_{xx}$ for the $\pm$chiral states at various magnetic fields at 250 K. In the whole magnetic field region, the data for the negative chirality almost overlap with those for the positive chirality, which indicates that the present phenomena do not depend on the sign of the chirality. In addition, the dependence
on the sign of the magnetic field is negligible in the whole magnetic field region. At 3.0 T, where the magnetic moments are forced to align ferromagnetically, $\Delta\rho_{xx}$ shows a gentle and smooth decrease as a function of bias current. On the other hand, all the $\Delta\rho_{xx}$ data below 2.5 T in the helimagnetic state show anomalies at some threshold bias current $j_\mathrm{th}$ in the helimagnetic state, similar to the 2.0 T case (\autoref{fig:FIG2}). The threshold bias current is almost minimum at 1.75 T. As the magnetic field is increased from 1.75 T, the anomalies in $\Delta\rho_{xx}$ become broad and the threshold current increases. 

In the lower magnetic field region, the bias current dependence of $\Delta\rho_{xx}$ becomes hysteretic and asymmetric. For example, at 0.0 T, when the bias current is increased from $-8.0\times10^9$ A/m$^2$, $\Delta\rho_{xx}$ does not show any distinct anomaly around $j_\mathrm{th}=-3.6\times10^9$ A/m$^2$, but does show a clear decrease at $j_\mathrm{th}=+3.6\times10^9$ A/m$^2$. On the other hand, when the bias current is decreased from $+8.0\times10^9$ A/m$^2$, $\Delta\rho_{xx}$ does not show any distinct anomaly at $j_\mathrm{th}=+3.6\times10^9$ A/m$^2$, but does show a clear increase at $j_\mathrm{th}=-3.6\times10^9$ A/m$^2$. The hysteretic behavior indicates that there are metastable states \cite{littlewood_SlidingCDW_numerical} below $j_\mathrm{th}$. As shown in the Supplemental Material \cite{supp}, this is not caused by the chirality degree of freedom. Because the hysteresis loop does not depend on the polarity of the magnetic field, this phenomenon is not related to any magnetic moment. Its origin will be discussed later. The magnitude of the threshold current decreases as the magnetic field increases from 0.0 T. Above 1.25 T, the magnitude of hysteresis and the asymmetry of bias current dependence decrease with increasing magnetic field. \hyperref[fig:FIG3]{Figure~\ref*{fig:FIG3}}(b) shows the magnetic field dependence of the threshold bias current. When the magnetic field is increased from 0 T, the threshold bias current first decreases and then shows a gentle upturn around 1.75 T. The threshold bias currents are almost symmetric with respect to the magnetic field and the bias current. A similar anomaly of $\Delta\rho_{xx}$ is widely observed below $T_c\approx340$ K, whereas it is absent above $T_c$. $j_\mathrm{th}$ increases with decreasing temperature \cite{supp}. 

Let us discuss the origin of the abrupt change of $\rho_{xx}$. As mentioned above, the bias current dependence of $\Delta\rho_{xx}$ is quite similar to the bias voltage dependence of resistivity in CDW/SDW systems \cite{gruner_CDW_transport,gruner_dynamics_SDW}. In addition, this phenomenon is observed only in the helimagnetic state. These indicate that the abrupt change of $\rho_{xx}$ is caused by the sliding motion of the helimagnetic structure. While the sliding motion is caused by a voltage in CDW/SDW systems, STT induced by the electric current should drive the sliding motion in the present case.
It is noted that the adiabatic spin-transfer torque should be dominant over Gilbert damping and nonadiabatic STT, whereas Gilbert damping or non-adiabatic torque is responsible for the chirality-control \cite{ohe_ChiralityControl}. When the chirality is not controlled, the domain walls between two chiral states seem to strongly pin the magnetic structure, and the onset electric current of sliding motion becomes larger than that in the chirality-controlled states. The decrease in resistivity can be explained by the nonadiabatic spin motive force \cite{duine_nonadiabaticSMF,tserkovnyak_SMF}. As a counteraction of STT, the dynamics of the magnetic moment induce a motive force \cite{volovik_linear_momentum,berger_SMF,barnes_SMF}. The adiabatic component should emerge as an inductive motive force in the imaginary part of resistivity \cite{yokouchi_emergent_inductor}, which seems to be small in this system. The nonadiabatic term is expressed as
\begin{equation}
    F_x=\beta (\partial_t \bm{m} \cdot\partial_x \bm{m}),
\end{equation}
where $\bm{m}$, $x$, $t$, and $\beta$ are the unit vector along the spatially dependent magnetic moment, position, time, and a constant, respectively. In the sliding state,
\begin{equation}
    \bm{m}=\big(\sin\theta, \cos\theta\sin\lambda k(x-vt), \cos\theta\cos\lambda k(x-vt)\big),
\end{equation}
where $\theta$, $\lambda=\pm1$, $k$, and $v$ are the conical angle, the chirality, the magnitude of the helical propagation vector, and the velocity of sliding motion, respectively. Then, the motive force in the sliding is\\
\begin{equation}
    F_x=-\lambda^2 k^2v\beta\cos^2\theta.
\end{equation}
According to a recent theoretical paper \cite{xie_sliding_SkXHL}, the velocity above the threshold is proportional to the electric current $j$ in the sliding state of a helimagnet. Therefore, a resistance decrease proportional to $F_x/j$ is expected by the nonadiabatic component of the spin motive force.

\begin{figure}[t]
\includegraphics{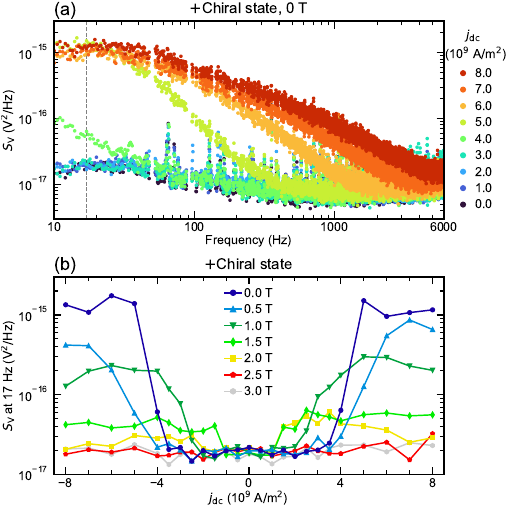}
\caption{(a) Voltage noise spectra under various $j_\mathrm{dc}$ for the $+$chiral state at 250 K and 0 T. (b) The voltage noise at 17~Hz as a function of $j_\mathrm{dc}$ under various magnetic fields at 250~K.}
\label{fig:FIG4}
\end{figure}

In the sliding state of CDW/SDW systems, in many cases, there is low-frequency fluctuation under an electric field above the threshold due to the interaction between the sliding density waves and pinning centers, which can be measured as voltage noise. We also measured the voltage noise of the helimagnetic $\mathrm{Mn}\mathrm{Au}_2$. To measure the voltage noise spectrum under a DC electric current, a spectrum analyzer (Rohde \& Schwarz, FSV 40) was connected to the voltage leads. Extrinsic spike-like noise insensitive to the electric current was omitted \cite{supp}. \hyperref[fig:FIG4]{Figure~\ref*{fig:FIG4}}(a) displays the noise spectra $S_\mathrm{V}$ for the $+$chiral state, represented in units of power spectral density ($V^2/\mathrm{Hz}$), under various bias electric currents at 250 K and 0 T. While the noise spectrum is suppressed below the threshold current $j_\mathrm{th}=3.6\times10^9$ A/m$^2$, a broad continuum emerges above $j_\mathrm{th}$. The emergence of a noise continuum in the low-frequency range, which is denoted as broadband noise (BBN), is widely observed when a periodic object such as CDW/SDW \cite{gruner_CDW_transport,gruner_dynamics_SDW,monceau_electronic}, magnetic texture\cite{sato_SkyrmionNoise, wang_singleSkNoise}, and vortex lattice in a superconductor \cite{marley_fluxNoise} shows incoherent translation (creep motion) under an electric current. The present observation of BBN supports the conclusion of sliding above $j_\mathrm{th}$ and suggests that the motion is the incoherent, creep type. The bias current dependence of noise magnitudes at 17 Hz in various magnetic fields for the $+$chiral state are shown in \autoref{fig:FIG4}(b). The noise magnitude abruptly increases at $j_\mathrm{th}$ in the low magnetic field region. The noise magnitude at 17 Hz as functions of $j_\mathrm{dc}$ and magnetic field is superimposed in \autoref{fig:FIG3}(b). It is clear from this figure that the noise magnitude is enhanced above $j_\mathrm{th}$, confirming that the BBN originates from the creep motion of the helimagnetic structure.

Finally, let us discuss the hysteresis and asymmetric shape of $\Delta\rho_{xx}$ in the low-field region and their magnetic field dependence. As mentioned above, $\Delta\rho_{xx}$ shows hysteresis in the low magnetic field region. Similar hystereses have been observed in incommensurate CDW/SDW \cite{gruner_CDW_transport,gruner_dynamics_SDW,monceau_electronic}, and the origin was ascribed to discommensurations. A discommensuration is a topological defect between two commensurate structures \cite{mcmillan_DCtheory} and is easily pinned by impurities \cite{rice_impurityDC,srolovitz_movingDC}. The distribution of discommensurations, depending on the history, may give rise to hysteresis. A similar mechanism may also have induced the hysteresis in this case, although more detailed study is needed to reach a final conclusion. Regarding the asymmetric $j_\mathrm{dc}$ dependence of $\Delta\rho_{xx}$, a literature suggested that unbalanced pinning due to contact electrodes caused asymmetric differential resistivity in a CDW system \cite{higgs_hysteresis}. Our samples were shaped into Hall bars with the use of a lithography technique, and no obvious scar was discerned \cite{supp}. Perhaps, some inevitable tiny and asymmetric roughness of the sample might cause the asymmetry of $\rho_{xx}$. In fact, the magnitude of hysteresis and asymmetry of $\rho_{xx}$ notably depend on the samples \cite{supp}.

In conclusion, we observed an abrupt change in differential resistivity at a threshold bias current in the helimagnetic state, which is ascribed to the sliding motion of the helimagnetic structure. This indicates that the low-energy collective excitation is similar to the CDW/SDW system, but the sliding motion of the helimagnetic structure is driven by an electric current, not by an electric field. Broadband voltage noise, which is commonly observed in various incoherent sliding (creeping) phenomena, emerged above the threshold current, confirming the sliding of the helimagnetic structure. A magnetic structure that can be manipulated with an electric current seems to be useful for the race track memories \cite{parkin_Racetrack,tomasello_SkLRacetrack}. In this sense, this work might pave the way to helimagnet-based spintronics.

\begin{acknowledgments}
We appreciate T. Sasaki for her help in the film deposition. The film deposition and device fabrication were carried out at the Cooperative Research and Development Center for Advanced Materials, IMR, Tohoku University. This work was supported by JSPS KAKENHI (Grant numbers JP20K03828, JP21H01036, JP22H04461, JP23H00232, JP23K13654, JP24H01638 and JP24H00189), JST SPRING(Grant No. JPMJSP2114), JST PRESTO (Grant No. JPMJPR19L6), and the Mitsubishi Foundation. Y.K. acknowledges support from GP-Spin at Tohoku University.
\end{acknowledgments}
\bibliography{MnAu2Sliding}

\providecommand{\noopsort}[1]{}\providecommand{\singleletter}[1]{#1}%
\begin{thebibliography}{47}%
\makeatletter
\providecommand \@ifxundefined [1]{%
 \@ifx{#1\undefined}
}%
\providecommand \@ifnum [1]{%
 \ifnum #1\expandafter \@firstoftwo
 \else \expandafter \@secondoftwo
 \fi
}%
\providecommand \@ifx [1]{%
 \ifx #1\expandafter \@firstoftwo
 \else \expandafter \@secondoftwo
 \fi
}%
\providecommand \natexlab [1]{#1}%
\providecommand \enquote  [1]{``#1''}%
\providecommand \bibnamefont  [1]{#1}%
\providecommand \bibfnamefont [1]{#1}%
\providecommand \citenamefont [1]{#1}%
\providecommand \href@noop [0]{\@secondoftwo}%
\providecommand \href [0]{\begingroup \@sanitize@url \@href}%
\providecommand \@href[1]{\@@startlink{#1}\@@href}%
\providecommand \@@href[1]{\endgroup#1\@@endlink}%
\providecommand \@sanitize@url [0]{\catcode `\\12\catcode `\$12\catcode `\&12\catcode `\#12\catcode `\^12\catcode `\_12\catcode `\%12\relax}%
\providecommand \@@startlink[1]{}%
\providecommand \@@endlink[0]{}%
\providecommand \url  [0]{\begingroup\@sanitize@url \@url }%
\providecommand \@url [1]{\endgroup\@href {#1}{\urlprefix }}%
\providecommand \urlprefix  [0]{URL }%
\providecommand \Eprint [0]{\href }%
\providecommand \doibase [0]{https://doi.org/}%
\providecommand \selectlanguage [0]{\@gobble}%
\providecommand \bibinfo  [0]{\@secondoftwo}%
\providecommand \bibfield  [0]{\@secondoftwo}%
\providecommand \translation [1]{[#1]}%
\providecommand \BibitemOpen [0]{}%
\providecommand \bibitemStop [0]{}%
\providecommand \bibitemNoStop [0]{.\EOS\space}%
\providecommand \EOS [0]{\spacefactor3000\relax}%
\providecommand \BibitemShut  [1]{\csname bibitem#1\endcsname}%
\let\auto@bib@innerbib\@empty
\bibitem [{\citenamefont {Grüner}\ and\ \citenamefont {Zettl}(1985)}]{gruner_CDW_transport}%
  \BibitemOpen
  \bibfield  {author} {\bibinfo {author} {\bibfnamefont {G.}~\bibnamefont {Grüner}}\ and\ \bibinfo {author} {\bibfnamefont {A.}~\bibnamefont {Zettl}},\ }\bibfield  {title} {\bibinfo {title} {Charge density wave conduction: {A} novel collective transport phenomenon in solids},\ }\href {https://doi.org/10.1016/0370-1573(85)90073-0} {\bibfield  {journal} {\bibinfo  {journal} {Phys. Rep.}\ }\textbf {\bibinfo {volume} {119}},\ \bibinfo {pages} {117} (\bibinfo {year} {1985})}\BibitemShut {NoStop}%
\bibitem [{\citenamefont {Grüner}(1994)}]{gruner_dynamics_SDW}%
  \BibitemOpen
  \bibfield  {author} {\bibinfo {author} {\bibfnamefont {G.}~\bibnamefont {Grüner}},\ }\bibfield  {title} {\bibinfo {title} {The dynamics of spin-density waves},\ }\href {https://doi.org/10.1103/RevModPhys.66.1} {\bibfield  {journal} {\bibinfo  {journal} {Rev. Mod. Phys.}\ }\textbf {\bibinfo {volume} {66}},\ \bibinfo {pages} {1} (\bibinfo {year} {1994})}\BibitemShut {NoStop}%
\bibitem [{\citenamefont {Monceau}(2012)}]{monceau_electronic}%
  \BibitemOpen
  \bibfield  {author} {\bibinfo {author} {\bibfnamefont {P.}~\bibnamefont {Monceau}},\ }\bibfield  {title} {\bibinfo {title} {Electronic crystals: an experimental overview},\ }\href {https://doi.org/10.1080/00018732.2012.719674} {\bibfield  {journal} {\bibinfo  {journal} {Adv. Phys.}\ }\textbf {\bibinfo {volume} {61}},\ \bibinfo {pages} {325} (\bibinfo {year} {2012})}\BibitemShut {NoStop}%
\bibitem [{\citenamefont {Lee}\ \emph {et~al.}(1974)\citenamefont {Lee}, \citenamefont {Rice},\ and\ \citenamefont {Anderson}}]{lee_conductivity_Sliding}%
  \BibitemOpen
  \bibfield  {author} {\bibinfo {author} {\bibfnamefont {P.~A.}\ \bibnamefont {Lee}}, \bibinfo {author} {\bibfnamefont {T.}~\bibnamefont {Rice}},\ and\ \bibinfo {author} {\bibfnamefont {P.~W.}\ \bibnamefont {Anderson}},\ }\bibfield  {title} {\bibinfo {title} {Conductivity from charge or spin density waves},\ }\href {https://www.sciencedirect.com/science/article/pii/0038109874908680} {\bibfield  {journal} {\bibinfo  {journal} {Solid State Commun.}\ }\textbf {\bibinfo {volume} {14}},\ \bibinfo {pages} {703} (\bibinfo {year} {1974})}\BibitemShut {NoStop}%
\bibitem [{\citenamefont {Shirahama}\ and\ \citenamefont {Kono}(1995)}]{shirahama_dynamical_LHe4}%
  \BibitemOpen
  \bibfield  {author} {\bibinfo {author} {\bibfnamefont {K.}~\bibnamefont {Shirahama}}\ and\ \bibinfo {author} {\bibfnamefont {K.}~\bibnamefont {Kono}},\ }\bibfield  {title} {\bibinfo {title} {Dynamical {Transition} in the {Wigner} {Solid} on a {Liquid} {Helium} {Surface}},\ }\href {https://doi.org/10.1103/PhysRevLett.74.781} {\bibfield  {journal} {\bibinfo  {journal} {Phys. Rev. Lett.}\ }\textbf {\bibinfo {volume} {74}},\ \bibinfo {pages} {781} (\bibinfo {year} {1995})}\BibitemShut {NoStop}%
\bibitem [{\citenamefont {Yamanouchi}\ \emph {et~al.}(1999)\citenamefont {Yamanouchi}, \citenamefont {Taguchi},\ and\ \citenamefont {Tokura}}]{yamanouchi_dielectric_sliding}%
  \BibitemOpen
  \bibfield  {author} {\bibinfo {author} {\bibfnamefont {S.}~\bibnamefont {Yamanouchi}}, \bibinfo {author} {\bibfnamefont {Y.}~\bibnamefont {Taguchi}},\ and\ \bibinfo {author} {\bibfnamefont {Y.}~\bibnamefont {Tokura}},\ }\bibfield  {title} {\bibinfo {title} {Dielectric {Breakdown} of the {Insulating} {Charge}-{Ordered} {State} in $\mathrm{La}_{2\ensuremath{-}x}\mathrm{Sr}_x\mathrm{Ni}\mathrm{O}_4$},\ }\href {https://doi.org/10.1103/PhysRevLett.83.5555} {\bibfield  {journal} {\bibinfo  {journal} {Phys. Rev. Lett.}\ }\textbf {\bibinfo {volume} {83}},\ \bibinfo {pages} {5555} (\bibinfo {year} {1999})}\BibitemShut {NoStop}%
\bibitem [{\citenamefont {Cox}\ \emph {et~al.}(2008)\citenamefont {Cox}, \citenamefont {Singleton}, \citenamefont {McDonald}, \citenamefont {Migliori},\ and\ \citenamefont {Littlewood}}]{cox_manganite_sliding}%
  \BibitemOpen
  \bibfield  {author} {\bibinfo {author} {\bibfnamefont {S.}~\bibnamefont {Cox}}, \bibinfo {author} {\bibfnamefont {J.}~\bibnamefont {Singleton}}, \bibinfo {author} {\bibfnamefont {R.~D.}\ \bibnamefont {McDonald}}, \bibinfo {author} {\bibfnamefont {A.}~\bibnamefont {Migliori}},\ and\ \bibinfo {author} {\bibfnamefont {P.~B.}\ \bibnamefont {Littlewood}},\ }\bibfield  {title} {\bibinfo {title} {Sliding charge-density wave in manganites},\ }\href {https://doi.org/10.1038/nmat2071} {\bibfield  {journal} {\bibinfo  {journal} {Nat. Mater.}\ }\textbf {\bibinfo {volume} {7}},\ \bibinfo {pages} {25} (\bibinfo {year} {2008})}\BibitemShut {NoStop}%
\bibitem [{\citenamefont {Blumberg}\ \emph {et~al.}(2002)\citenamefont {Blumberg}, \citenamefont {Littlewood}, \citenamefont {Gozar}, \citenamefont {Dennis}, \citenamefont {Motoyama}, \citenamefont {Eisaki},\ and\ \citenamefont {Uchida}}]{blumberg_SCOsliding}%
  \BibitemOpen
  \bibfield  {author} {\bibinfo {author} {\bibfnamefont {G.}~\bibnamefont {Blumberg}}, \bibinfo {author} {\bibfnamefont {P.}~\bibnamefont {Littlewood}}, \bibinfo {author} {\bibfnamefont {A.}~\bibnamefont {Gozar}}, \bibinfo {author} {\bibfnamefont {B.~S.}\ \bibnamefont {Dennis}}, \bibinfo {author} {\bibfnamefont {N.}~\bibnamefont {Motoyama}}, \bibinfo {author} {\bibfnamefont {H.}~\bibnamefont {Eisaki}},\ and\ \bibinfo {author} {\bibfnamefont {S.}~\bibnamefont {Uchida}},\ }\bibfield  {title} {\bibinfo {title} {Sliding {Density} {Wave} in $\mathrm{Sr}_{14}\mathrm{Cu}_{24}\mathrm{O}_{41}$ {Ladder} {Compounds}},\ }\href {https://doi.org/10.1126/science.1070481} {\bibfield  {journal} {\bibinfo  {journal} {Science}\ }\textbf {\bibinfo {volume} {297}},\ \bibinfo {pages} {584} (\bibinfo {year} {2002})}\BibitemShut {NoStop}%
\bibitem [{\citenamefont {Brataas}\ \emph {et~al.}(2012)\citenamefont {Brataas}, \citenamefont {Kent},\ and\ \citenamefont {Ohno}}]{brataas_current-induced_torque}%
  \BibitemOpen
  \bibfield  {author} {\bibinfo {author} {\bibfnamefont {A.}~\bibnamefont {Brataas}}, \bibinfo {author} {\bibfnamefont {A.~D.}\ \bibnamefont {Kent}},\ and\ \bibinfo {author} {\bibfnamefont {H.}~\bibnamefont {Ohno}},\ }\bibfield  {title} {\bibinfo {title} {Current-induced torques in magnetic materials},\ }\href {https://doi.org/10.1038/nmat3311} {\bibfield  {journal} {\bibinfo  {journal} {Nat. Mater.}\ }\textbf {\bibinfo {volume} {11}},\ \bibinfo {pages} {372} (\bibinfo {year} {2012})}\BibitemShut {NoStop}%
\bibitem [{\citenamefont {Yamaguchi}\ \emph {et~al.}(2004)\citenamefont {Yamaguchi}, \citenamefont {Ono}, \citenamefont {Nasu}, \citenamefont {Miyake}, \citenamefont {Mibu},\ and\ \citenamefont {Shinjo}}]{yamaguchi_FMDW_Observation}%
  \BibitemOpen
  \bibfield  {author} {\bibinfo {author} {\bibfnamefont {A.}~\bibnamefont {Yamaguchi}}, \bibinfo {author} {\bibfnamefont {T.}~\bibnamefont {Ono}}, \bibinfo {author} {\bibfnamefont {S.}~\bibnamefont {Nasu}}, \bibinfo {author} {\bibfnamefont {K.}~\bibnamefont {Miyake}}, \bibinfo {author} {\bibfnamefont {K.}~\bibnamefont {Mibu}},\ and\ \bibinfo {author} {\bibfnamefont {T.}~\bibnamefont {Shinjo}},\ }\bibfield  {title} {\bibinfo {title} {Real-{Space} {Observation} of {Current}-{Driven} {Domain} {Wall} {Motion} in {Submicron} {Magnetic} {Wires}},\ }\href {https://doi.org/10.1103/PhysRevLett.92.077205} {\bibfield  {journal} {\bibinfo  {journal} {Phys. Rev. Lett.}\ }\textbf {\bibinfo {volume} {92}},\ \bibinfo {pages} {077205} (\bibinfo {year} {2004})}\BibitemShut {NoStop}%
\bibitem [{\citenamefont {Koyama}\ \emph {et~al.}(2011)\citenamefont {Koyama}, \citenamefont {Chiba}, \citenamefont {Ueda}, \citenamefont {Kondou}, \citenamefont {Tanigawa}, \citenamefont {Fukami}, \citenamefont {Suzuki}, \citenamefont {Ohshima}, \citenamefont {Ishiwata}, \citenamefont {Nakatani}, \citenamefont {Kobayashi},\ and\ \citenamefont {Ono}}]{koyama_intrinsic_pinning}%
  \BibitemOpen
  \bibfield  {author} {\bibinfo {author} {\bibfnamefont {T.}~\bibnamefont {Koyama}}, \bibinfo {author} {\bibfnamefont {D.}~\bibnamefont {Chiba}}, \bibinfo {author} {\bibfnamefont {K.}~\bibnamefont {Ueda}}, \bibinfo {author} {\bibfnamefont {K.}~\bibnamefont {Kondou}}, \bibinfo {author} {\bibfnamefont {H.}~\bibnamefont {Tanigawa}}, \bibinfo {author} {\bibfnamefont {S.}~\bibnamefont {Fukami}}, \bibinfo {author} {\bibfnamefont {T.}~\bibnamefont {Suzuki}}, \bibinfo {author} {\bibfnamefont {N.}~\bibnamefont {Ohshima}}, \bibinfo {author} {\bibfnamefont {N.}~\bibnamefont {Ishiwata}}, \bibinfo {author} {\bibfnamefont {Y.}~\bibnamefont {Nakatani}}, \bibinfo {author} {\bibfnamefont {K.}~\bibnamefont {Kobayashi}},\ and\ \bibinfo {author} {\bibfnamefont {T.}~\bibnamefont {Ono}},\ }\bibfield  {title} {\bibinfo {title} {Observation of the intrinsic pinning of a magnetic domain wall in a ferromagnetic nanowire},\ }\href {https://doi.org/10.1038/nmat2961} {\bibfield  {journal} {\bibinfo  {journal} {Nat. Mater.}\ }\textbf
  {\bibinfo {volume} {10}},\ \bibinfo {pages} {194} (\bibinfo {year} {2011})}\BibitemShut {NoStop}%
\bibitem [{\citenamefont {Yamanouchi}\ \emph {et~al.}(2004)\citenamefont {Yamanouchi}, \citenamefont {Chiba}, \citenamefont {Matsukura},\ and\ \citenamefont {Ohno}}]{yamanouchi_current-induced_GaMnAsDW}%
  \BibitemOpen
  \bibfield  {author} {\bibinfo {author} {\bibfnamefont {M.}~\bibnamefont {Yamanouchi}}, \bibinfo {author} {\bibfnamefont {D.}~\bibnamefont {Chiba}}, \bibinfo {author} {\bibfnamefont {F.}~\bibnamefont {Matsukura}},\ and\ \bibinfo {author} {\bibfnamefont {H.}~\bibnamefont {Ohno}},\ }\bibfield  {title} {\bibinfo {title} {Current-induced domain-wall switching in a ferromagnetic semiconductor structure},\ }\href {https://doi.org/10.1038/nature02441} {\bibfield  {journal} {\bibinfo  {journal} {Nature}\ }\textbf {\bibinfo {volume} {428}},\ \bibinfo {pages} {539} (\bibinfo {year} {2004})}\BibitemShut {NoStop}%
\bibitem [{\citenamefont {Jonietz}\ \emph {et~al.}(2010)\citenamefont {Jonietz}, \citenamefont {Mühlbauer}, \citenamefont {Pfleiderer}, \citenamefont {Neubauer}, \citenamefont {Münzer}, \citenamefont {Bauer}, \citenamefont {Adams}, \citenamefont {Georgii}, \citenamefont {Böni}, \citenamefont {Duine}, \citenamefont {Everschor}, \citenamefont {Garst},\ and\ \citenamefont {Rosch}}]{jonietz_STT_MnSi}%
  \BibitemOpen
  \bibfield  {author} {\bibinfo {author} {\bibfnamefont {F.}~\bibnamefont {Jonietz}}, \bibinfo {author} {\bibfnamefont {S.}~\bibnamefont {Mühlbauer}}, \bibinfo {author} {\bibfnamefont {C.}~\bibnamefont {Pfleiderer}}, \bibinfo {author} {\bibfnamefont {A.}~\bibnamefont {Neubauer}}, \bibinfo {author} {\bibfnamefont {W.}~\bibnamefont {Münzer}}, \bibinfo {author} {\bibfnamefont {A.}~\bibnamefont {Bauer}}, \bibinfo {author} {\bibfnamefont {T.}~\bibnamefont {Adams}}, \bibinfo {author} {\bibfnamefont {R.}~\bibnamefont {Georgii}}, \bibinfo {author} {\bibfnamefont {P.}~\bibnamefont {Böni}}, \bibinfo {author} {\bibfnamefont {R.~A.}\ \bibnamefont {Duine}}, \bibinfo {author} {\bibfnamefont {K.}~\bibnamefont {Everschor}}, \bibinfo {author} {\bibfnamefont {M.}~\bibnamefont {Garst}},\ and\ \bibinfo {author} {\bibfnamefont {A.}~\bibnamefont {Rosch}},\ }\bibfield  {title} {\bibinfo {title} {Spin {Transfer} {Torques} in {MnSi} at {Ultralow} {Current} {Densities}},\ }\href {https://doi.org/10.1126/science.1195709} {\bibfield
  {journal} {\bibinfo  {journal} {Science}\ }\textbf {\bibinfo {volume} {330}},\ \bibinfo {pages} {1648} (\bibinfo {year} {2010})}\BibitemShut {NoStop}%
\bibitem [{\citenamefont {Schulz}\ \emph {et~al.}(2012)\citenamefont {Schulz}, \citenamefont {Ritz}, \citenamefont {Bauer}, \citenamefont {Halder}, \citenamefont {Wagner}, \citenamefont {Franz}, \citenamefont {Pfleiderer}, \citenamefont {Everschor}, \citenamefont {Garst},\ and\ \citenamefont {Rosch}}]{schulz_emergent_skyrmion}%
  \BibitemOpen
  \bibfield  {author} {\bibinfo {author} {\bibfnamefont {T.}~\bibnamefont {Schulz}}, \bibinfo {author} {\bibfnamefont {R.}~\bibnamefont {Ritz}}, \bibinfo {author} {\bibfnamefont {A.}~\bibnamefont {Bauer}}, \bibinfo {author} {\bibfnamefont {M.}~\bibnamefont {Halder}}, \bibinfo {author} {\bibfnamefont {M.}~\bibnamefont {Wagner}}, \bibinfo {author} {\bibfnamefont {C.}~\bibnamefont {Franz}}, \bibinfo {author} {\bibfnamefont {C.}~\bibnamefont {Pfleiderer}}, \bibinfo {author} {\bibfnamefont {K.}~\bibnamefont {Everschor}}, \bibinfo {author} {\bibfnamefont {M.}~\bibnamefont {Garst}},\ and\ \bibinfo {author} {\bibfnamefont {A.}~\bibnamefont {Rosch}},\ }\bibfield  {title} {\bibinfo {title} {Emergent electrodynamics of skyrmions in a chiral magnet},\ }\href {https://doi.org/10.1038/nphys2231} {\bibfield  {journal} {\bibinfo  {journal} {Nat. Phys.}\ }\textbf {\bibinfo {volume} {8}},\ \bibinfo {pages} {301} (\bibinfo {year} {2012})}\BibitemShut {NoStop}%
\bibitem [{\citenamefont {Yokouchi}\ \emph {et~al.}(2018)\citenamefont {Yokouchi}, \citenamefont {Hoshino}, \citenamefont {Kanazawa}, \citenamefont {Kikkawa}, \citenamefont {Morikawa}, \citenamefont {Shibata}, \citenamefont {Arima}, \citenamefont {Taguchi}, \citenamefont {Kagawa}, \citenamefont {Nagaosa},\ and\ \citenamefont {Tokura}}]{yokouchi_current-induced_skyrmion}%
  \BibitemOpen
  \bibfield  {author} {\bibinfo {author} {\bibfnamefont {T.}~\bibnamefont {Yokouchi}}, \bibinfo {author} {\bibfnamefont {S.}~\bibnamefont {Hoshino}}, \bibinfo {author} {\bibfnamefont {N.}~\bibnamefont {Kanazawa}}, \bibinfo {author} {\bibfnamefont {A.}~\bibnamefont {Kikkawa}}, \bibinfo {author} {\bibfnamefont {D.}~\bibnamefont {Morikawa}}, \bibinfo {author} {\bibfnamefont {K.}~\bibnamefont {Shibata}}, \bibinfo {author} {\bibfnamefont {T.}~\bibnamefont {Arima}}, \bibinfo {author} {\bibfnamefont {Y.}~\bibnamefont {Taguchi}}, \bibinfo {author} {\bibfnamefont {F.}~\bibnamefont {Kagawa}}, \bibinfo {author} {\bibfnamefont {N.}~\bibnamefont {Nagaosa}},\ and\ \bibinfo {author} {\bibfnamefont {Y.}~\bibnamefont {Tokura}},\ }\bibfield  {title} {\bibinfo {title} {Current-induced dynamics of skyrmion strings},\ }\href {https://doi.org/10.1126/sciadv.aat1115} {\bibfield  {journal} {\bibinfo  {journal} {Sci. Adv.}\ }\textbf {\bibinfo {volume} {4}},\ \bibinfo {pages} {eaat1115} (\bibinfo {year} {2018})}\BibitemShut {NoStop}%
\bibitem [{\citenamefont {Sato}\ \emph {et~al.}(2019)\citenamefont {Sato}, \citenamefont {Koshibae}, \citenamefont {Kikkawa}, \citenamefont {Yokouchi}, \citenamefont {Oike}, \citenamefont {Taguchi}, \citenamefont {Nagaosa}, \citenamefont {Tokura},\ and\ \citenamefont {Kagawa}}]{sato_SkyrmionNoise}%
  \BibitemOpen
  \bibfield  {author} {\bibinfo {author} {\bibfnamefont {T.}~\bibnamefont {Sato}}, \bibinfo {author} {\bibfnamefont {W.}~\bibnamefont {Koshibae}}, \bibinfo {author} {\bibfnamefont {A.}~\bibnamefont {Kikkawa}}, \bibinfo {author} {\bibfnamefont {T.}~\bibnamefont {Yokouchi}}, \bibinfo {author} {\bibfnamefont {H.}~\bibnamefont {Oike}}, \bibinfo {author} {\bibfnamefont {Y.}~\bibnamefont {Taguchi}}, \bibinfo {author} {\bibfnamefont {N.}~\bibnamefont {Nagaosa}}, \bibinfo {author} {\bibfnamefont {Y.}~\bibnamefont {Tokura}},\ and\ \bibinfo {author} {\bibfnamefont {F.}~\bibnamefont {Kagawa}},\ }\bibfield  {title} {\bibinfo {title} {Slow steady flow of a skyrmion lattice in a confined geometry probed by narrow-band resistance noise},\ }\href {https://doi.org/10.1103/PhysRevB.100.094410} {\bibfield  {journal} {\bibinfo  {journal} {Phys. Rev. B}\ }\textbf {\bibinfo {volume} {100}},\ \bibinfo {pages} {094410} (\bibinfo {year} {2019})}\BibitemShut {NoStop}%
\bibitem [{\citenamefont {He}\ \emph {et~al.}(2024)\citenamefont {He}, \citenamefont {Li}, \citenamefont {Chen}, \citenamefont {Wang}, \citenamefont {Shen}, \citenamefont {Wang}, \citenamefont {Song}, \citenamefont {Zhao}, \citenamefont {Cai}, \citenamefont {Lin}, \citenamefont {Zhang},\ and\ \citenamefont {Shen}}]{he_antiSKy_sliding}%
  \BibitemOpen
  \bibfield  {author} {\bibinfo {author} {\bibfnamefont {Z.}~\bibnamefont {He}}, \bibinfo {author} {\bibfnamefont {Z.}~\bibnamefont {Li}}, \bibinfo {author} {\bibfnamefont {Z.}~\bibnamefont {Chen}}, \bibinfo {author} {\bibfnamefont {Z.}~\bibnamefont {Wang}}, \bibinfo {author} {\bibfnamefont {J.}~\bibnamefont {Shen}}, \bibinfo {author} {\bibfnamefont {S.}~\bibnamefont {Wang}}, \bibinfo {author} {\bibfnamefont {C.}~\bibnamefont {Song}}, \bibinfo {author} {\bibfnamefont {T.}~\bibnamefont {Zhao}}, \bibinfo {author} {\bibfnamefont {J.}~\bibnamefont {Cai}}, \bibinfo {author} {\bibfnamefont {S.-Z.}\ \bibnamefont {Lin}}, \bibinfo {author} {\bibfnamefont {Y.}~\bibnamefont {Zhang}},\ and\ \bibinfo {author} {\bibfnamefont {B.}~\bibnamefont {Shen}},\ }\bibfield  {title} {\bibinfo {title} {Experimental observation of current-driven antiskyrmion sliding in stripe domains},\ }\href {https://doi.org/10.1038/s41563-024-01870-8} {\bibfield  {journal} {\bibinfo  {journal} {Nat. Mater.}\ }\textbf {\bibinfo {volume} {23}},\
  \bibinfo {pages} {1048} (\bibinfo {year} {2024})}\BibitemShut {NoStop}%
\bibitem [{\citenamefont {Birch}\ \emph {et~al.}(2024)\citenamefont {Birch}, \citenamefont {Belopolski}, \citenamefont {Fujishiro}, \citenamefont {Kawamura}, \citenamefont {Kikkawa}, \citenamefont {Taguchi}, \citenamefont {Hirschberger}, \citenamefont {Nagaosa},\ and\ \citenamefont {Tokura}}]{birch_Sky_sliding}%
  \BibitemOpen
  \bibfield  {author} {\bibinfo {author} {\bibfnamefont {M.~T.}\ \bibnamefont {Birch}}, \bibinfo {author} {\bibfnamefont {I.}~\bibnamefont {Belopolski}}, \bibinfo {author} {\bibfnamefont {Y.}~\bibnamefont {Fujishiro}}, \bibinfo {author} {\bibfnamefont {M.}~\bibnamefont {Kawamura}}, \bibinfo {author} {\bibfnamefont {A.}~\bibnamefont {Kikkawa}}, \bibinfo {author} {\bibfnamefont {Y.}~\bibnamefont {Taguchi}}, \bibinfo {author} {\bibfnamefont {M.}~\bibnamefont {Hirschberger}}, \bibinfo {author} {\bibfnamefont {N.}~\bibnamefont {Nagaosa}},\ and\ \bibinfo {author} {\bibfnamefont {Y.}~\bibnamefont {Tokura}},\ }\bibfield  {title} {\bibinfo {title} {Dynamic transition and {Galilean} relativity of current-driven skyrmions},\ }\href {https://doi.org/10.1038/s41586-024-07859-2} {\bibfield  {journal} {\bibinfo  {journal} {Nature}\ }\textbf {\bibinfo {volume} {633}},\ \bibinfo {pages} {554} (\bibinfo {year} {2024})}\BibitemShut {NoStop}%
\bibitem [{\citenamefont {Tserkovnyak}\ and\ \citenamefont {Brataas}(2005)}]{tserkovnyak_adiabatic_pumping}%
  \BibitemOpen
  \bibfield  {author} {\bibinfo {author} {\bibfnamefont {Y.}~\bibnamefont {Tserkovnyak}}\ and\ \bibinfo {author} {\bibfnamefont {A.}~\bibnamefont {Brataas}},\ }\bibfield  {title} {\bibinfo {title} {Spontaneous-symmetry-breaking mechanism of adiabatic pumping},\ }\href {https://doi.org/10.1103/PhysRevB.71.052406} {\bibfield  {journal} {\bibinfo  {journal} {Phys. Rev. B}\ }\textbf {\bibinfo {volume} {71}},\ \bibinfo {pages} {052406} (\bibinfo {year} {2005})}\BibitemShut {NoStop}%
\bibitem [{\citenamefont {Wessely}\ \emph {et~al.}(2006)\citenamefont {Wessely}, \citenamefont {Skubic},\ and\ \citenamefont {Nordstrom}}]{wessely_current_dynamics_helimag}%
  \BibitemOpen
  \bibfield  {author} {\bibinfo {author} {\bibfnamefont {O.}~\bibnamefont {Wessely}}, \bibinfo {author} {\bibfnamefont {B.}~\bibnamefont {Skubic}},\ and\ \bibinfo {author} {\bibfnamefont {L.}~\bibnamefont {Nordstrom}},\ }\bibfield  {title} {\bibinfo {title} {Current {Driven} {Magnetization} {Dynamics} in {Helical} {Spin} {Density} {Waves}},\ }\href {https://doi.org/10.1103/PhysRevLett.96.256601} {\bibfield  {journal} {\bibinfo  {journal} {Phys. Rev. Lett.}\ }\textbf {\bibinfo {volume} {96}},\ \bibinfo {pages} {256601} (\bibinfo {year} {2006})}\BibitemShut {NoStop}%
\bibitem [{\citenamefont {Wessely}\ \emph {et~al.}(2009)\citenamefont {Wessely}, \citenamefont {Skubic},\ and\ \citenamefont {Nordstrom}}]{wessely_STT_helical}%
  \BibitemOpen
  \bibfield  {author} {\bibinfo {author} {\bibfnamefont {O.}~\bibnamefont {Wessely}}, \bibinfo {author} {\bibfnamefont {B.}~\bibnamefont {Skubic}},\ and\ \bibinfo {author} {\bibfnamefont {L.}~\bibnamefont {Nordstrom}},\ }\bibfield  {title} {\bibinfo {title} {Spin-transfer torque in helical spin-density waves},\ }\href {https://doi.org/10.1103/PhysRevB.79.104433} {\bibfield  {journal} {\bibinfo  {journal} {Phys. Rev. B}\ }\textbf {\bibinfo {volume} {79}},\ \bibinfo {pages} {104433} (\bibinfo {year} {2009})}\BibitemShut {NoStop}%
\bibitem [{\citenamefont {Iwasaki}\ \emph {et~al.}(2013)\citenamefont {Iwasaki}, \citenamefont {Mochizuki},\ and\ \citenamefont {Nagaosa}}]{iwasaki_universal_HLSk}%
  \BibitemOpen
  \bibfield  {author} {\bibinfo {author} {\bibfnamefont {J.}~\bibnamefont {Iwasaki}}, \bibinfo {author} {\bibfnamefont {M.}~\bibnamefont {Mochizuki}},\ and\ \bibinfo {author} {\bibfnamefont {N.}~\bibnamefont {Nagaosa}},\ }\bibfield  {title} {\bibinfo {title} {Universal current-velocity relation of skyrmion motion in chiral magnets},\ }\href {https://doi.org/10.1038/ncomms2442} {\bibfield  {journal} {\bibinfo  {journal} {Nat. Commun.}\ }\textbf {\bibinfo {volume} {4}},\ \bibinfo {pages} {1463} (\bibinfo {year} {2013})}\BibitemShut {NoStop}%
\bibitem [{\citenamefont {Del~Ser}\ \emph {et~al.}(2021)\citenamefont {Del~Ser}, \citenamefont {Heinen},\ and\ \citenamefont {Rosch}}]{del_archimedean}%
  \BibitemOpen
  \bibfield  {author} {\bibinfo {author} {\bibfnamefont {N.}~\bibnamefont {Del~Ser}}, \bibinfo {author} {\bibfnamefont {L.}~\bibnamefont {Heinen}},\ and\ \bibinfo {author} {\bibfnamefont {A.}~\bibnamefont {Rosch}},\ }\bibfield  {title} {\bibinfo {title} {Archimedean screw in driven chiral magnets},\ }\href {https://doi.org/10.21468/SciPostPhys.11.1.009} {\bibfield  {journal} {\bibinfo  {journal} {SciPost Phys.}\ }\textbf {\bibinfo {volume} {11}},\ \bibinfo {pages} {009} (\bibinfo {year} {2021})}\BibitemShut {NoStop}%
\bibitem [{\citenamefont {Ustinov}\ and\ \citenamefont {Yasyulevich}(2022)}]{ustinov_current-induced_rotation_helimag}%
  \BibitemOpen
  \bibfield  {author} {\bibinfo {author} {\bibfnamefont {V.~V.}\ \bibnamefont {Ustinov}}\ and\ \bibinfo {author} {\bibfnamefont {I.~A.}\ \bibnamefont {Yasyulevich}},\ }\bibfield  {title} {\bibinfo {title} {Chirality-dependent spin-transfer torque and current-induced spin rotation in helimagnets},\ }\href {https://doi.org/10.1103/PhysRevB.106.064417} {\bibfield  {journal} {\bibinfo  {journal} {Phys. Rev. B}\ }\textbf {\bibinfo {volume} {106}},\ \bibinfo {pages} {064417} (\bibinfo {year} {2022})}\BibitemShut {NoStop}%
\bibitem [{\citenamefont {Kurebayashi}\ \emph {et~al.}(2022)\citenamefont {Kurebayashi}, \citenamefont {Liu}, \citenamefont {Masell},\ and\ \citenamefont {Nagaosa}}]{kurebayashi_pumping}%
  \BibitemOpen
  \bibfield  {author} {\bibinfo {author} {\bibfnamefont {D.}~\bibnamefont {Kurebayashi}}, \bibinfo {author} {\bibfnamefont {Y.}~\bibnamefont {Liu}}, \bibinfo {author} {\bibfnamefont {J.}~\bibnamefont {Masell}},\ and\ \bibinfo {author} {\bibfnamefont {N.}~\bibnamefont {Nagaosa}},\ }\bibfield  {title} {\bibinfo {title} {Theory of charge and spin pumping in atomic-scale spiral magnets},\ }\href {https://doi.org/10.1103/PhysRevB.106.205110} {\bibfield  {journal} {\bibinfo  {journal} {Phys. Rev. B}\ }\textbf {\bibinfo {volume} {106}},\ \bibinfo {pages} {205110} (\bibinfo {year} {2022})}\BibitemShut {NoStop}%
\bibitem [{\citenamefont {Xie}\ \emph {et~al.}(2024)\citenamefont {Xie}, \citenamefont {Liu},\ and\ \citenamefont {Nagaosa}}]{xie_sliding_SkXHL}%
  \BibitemOpen
  \bibfield  {author} {\bibinfo {author} {\bibfnamefont {Y.-M.}\ \bibnamefont {Xie}}, \bibinfo {author} {\bibfnamefont {Y.}~\bibnamefont {Liu}},\ and\ \bibinfo {author} {\bibfnamefont {N.}~\bibnamefont {Nagaosa}},\ }\bibfield  {title} {\bibinfo {title} {Sliding {Dynamics} of {Current}-{Driven} {Skyrmion} {Crystal} and {Helix} in {Chiral} {Magnets}},\ }\href {https://doi.org/10.1103/PhysRevLett.133.096702} {\bibfield  {journal} {\bibinfo  {journal} {Phys. Rev. Lett.}\ }\textbf {\bibinfo {volume} {133}},\ \bibinfo {pages} {096702} (\bibinfo {year} {2024})}\BibitemShut {NoStop}%
\bibitem [{\citenamefont {Hall}\ and\ \citenamefont {Royan}(1959)}]{hall_MnAu2structure}%
  \BibitemOpen
  \bibfield  {author} {\bibinfo {author} {\bibfnamefont {E.~O.}\ \bibnamefont {Hall}}\ and\ \bibinfo {author} {\bibfnamefont {J.}~\bibnamefont {Royan}},\ }\bibfield  {title} {\bibinfo {title} {The structure of $\mathrm{Au}_2\mathrm{Mn}$},\ }\href {https://doi.org/10.1107/S0365110X59001761} {\bibfield  {journal} {\bibinfo  {journal} {Acta Cryst.}\ }\textbf {\bibinfo {volume} {12}},\ \bibinfo {pages} {607} (\bibinfo {year} {1959})}\BibitemShut {NoStop}%
\bibitem [{\citenamefont {Herpin}\ and\ \citenamefont {Meriel}(1961)}]{herpin_MnAu2neutron}%
  \BibitemOpen
  \bibfield  {author} {\bibinfo {author} {\bibfnamefont {A.}~\bibnamefont {Herpin}}\ and\ \bibinfo {author} {\bibfnamefont {P.}~\bibnamefont {Meriel}},\ }\bibfield  {title} {\bibinfo {title} {Étude de l'antiferromagnétisme helicoidal de {MnAu2} par diffraction de neutrons},\ }\href {https://doi.org/10.1051/jphysrad:01961002206033700} {\bibfield  {journal} {\bibinfo  {journal} {J. Phys. Radium}\ }\textbf {\bibinfo {volume} {22}},\ \bibinfo {pages} {337} (\bibinfo {year} {1961})}\BibitemShut {NoStop}%
\bibitem [{\citenamefont {Meyer}\ and\ \citenamefont {Taglang}(1956)}]{meyer_MnAu2Tc}%
  \BibitemOpen
  \bibfield  {author} {\bibinfo {author} {\bibfnamefont {A.~J.}\ \bibnamefont {Meyer}}\ and\ \bibinfo {author} {\bibfnamefont {P.}~\bibnamefont {Taglang}},\ }\bibfield  {title} {\bibinfo {title} {Propriétés magnétiques, antiferromagnétisme et ferromagnétisme de {MnAu} 2},\ }\href {https://doi.org/10.1051/jphysrad:01956001706045700} {\bibfield  {journal} {\bibinfo  {journal} {J. Phys. Radium}\ }\textbf {\bibinfo {volume} {17}},\ \bibinfo {pages} {457} (\bibinfo {year} {1956})}\BibitemShut {NoStop}%
\bibitem [{\citenamefont {Masuda}\ \emph {et~al.}(2024)\citenamefont {Masuda}, \citenamefont {Seki}, \citenamefont {Ohe}, \citenamefont {Nii}, \citenamefont {Masuda}, \citenamefont {Takanashi},\ and\ \citenamefont {Onose}}]{masuda_chirality_control_MnAu2}%
  \BibitemOpen
  \bibfield  {author} {\bibinfo {author} {\bibfnamefont {H.}~\bibnamefont {Masuda}}, \bibinfo {author} {\bibfnamefont {T.}~\bibnamefont {Seki}}, \bibinfo {author} {\bibfnamefont {J.}~\bibnamefont {Ohe}}, \bibinfo {author} {\bibfnamefont {Y.}~\bibnamefont {Nii}}, \bibinfo {author} {\bibfnamefont {H.}~\bibnamefont {Masuda}}, \bibinfo {author} {\bibfnamefont {K.}~\bibnamefont {Takanashi}},\ and\ \bibinfo {author} {\bibfnamefont {Y.}~\bibnamefont {Onose}},\ }\bibfield  {title} {\bibinfo {title} {Room temperature chirality switching and detection in a helimagnetic $\mathrm{Mn}\mathrm{Au}_2$ thin film},\ }\href {https://doi.org/10.1038/s41467-024-46326-4} {\bibfield  {journal} {\bibinfo  {journal} {Nat. Commun.}\ }\textbf {\bibinfo {volume} {15}},\ \bibinfo {pages} {1999} (\bibinfo {year} {2024})}\BibitemShut {NoStop}%
\bibitem [{sup()}]{supp}%
  \BibitemOpen
  \href@noop {} {}\bibinfo {note} {See Supplemental Material at [URL will be inserted by publisher] for details on the analysis of transition temperature and voltage noise spectra, as well as additional data on $\rho_{xx}$, voltage noise, and nonreciprocal electronic transport.}\BibitemShut {Stop}%
\bibitem [{\citenamefont {Littlewood}(1986)}]{littlewood_SlidingCDW_numerical}%
  \BibitemOpen
  \bibfield  {author} {\bibinfo {author} {\bibfnamefont {P.~B.}\ \bibnamefont {Littlewood}},\ }\bibfield  {title} {\bibinfo {title} {Sliding charge-density waves: {A} numerical study},\ }\href {https://doi.org/10.1103/PhysRevB.33.6694} {\bibfield  {journal} {\bibinfo  {journal} {Phys. Rev. B}\ }\textbf {\bibinfo {volume} {33}},\ \bibinfo {pages} {6694} (\bibinfo {year} {1986})}\BibitemShut {NoStop}%
\bibitem [{\citenamefont {Ohe}\ and\ \citenamefont {Onose}(2021)}]{ohe_ChiralityControl}%
  \BibitemOpen
  \bibfield  {author} {\bibinfo {author} {\bibfnamefont {J.}~\bibnamefont {Ohe}}\ and\ \bibinfo {author} {\bibfnamefont {Y.}~\bibnamefont {Onose}},\ }\bibfield  {title} {\bibinfo {title} {Chirality control of the spin structure in monoaxial helimagnets by charge current},\ }\href {https://doi.org/10.1063/5.0037357} {\bibfield  {journal} {\bibinfo  {journal} {Appl. Phys. Lett.}\ }\textbf {\bibinfo {volume} {118}},\ \bibinfo {pages} {042407} (\bibinfo {year} {2021})}\BibitemShut {NoStop}%
\bibitem [{\citenamefont {Duine}(2008)}]{duine_nonadiabaticSMF}%
  \BibitemOpen
  \bibfield  {author} {\bibinfo {author} {\bibfnamefont {R.~A.}\ \bibnamefont {Duine}},\ }\bibfield  {title} {\bibinfo {title} {Spin pumping by a field-driven domain wall},\ }\href {https://doi.org/10.1103/PhysRevB.77.014409} {\bibfield  {journal} {\bibinfo  {journal} {Phys. Rev. B}\ }\textbf {\bibinfo {volume} {77}},\ \bibinfo {pages} {014409} (\bibinfo {year} {2008})}\BibitemShut {NoStop}%
\bibitem [{\citenamefont {Tserkovnyak}\ and\ \citenamefont {Mecklenburg}(2008)}]{tserkovnyak_SMF}%
  \BibitemOpen
  \bibfield  {author} {\bibinfo {author} {\bibfnamefont {Y.}~\bibnamefont {Tserkovnyak}}\ and\ \bibinfo {author} {\bibfnamefont {M.}~\bibnamefont {Mecklenburg}},\ }\bibfield  {title} {\bibinfo {title} {Electron transport driven by nonequilibrium magnetic textures},\ }\href {https://doi.org/10.1103/PhysRevB.77.134407} {\bibfield  {journal} {\bibinfo  {journal} {Phys. Rev. B}\ }\textbf {\bibinfo {volume} {77}},\ \bibinfo {pages} {134407} (\bibinfo {year} {2008})}\BibitemShut {NoStop}%
\bibitem [{\citenamefont {Volovik}(1987)}]{volovik_linear_momentum}%
  \BibitemOpen
  \bibfield  {author} {\bibinfo {author} {\bibfnamefont {G.~E.}\ \bibnamefont {Volovik}},\ }\bibfield  {title} {\bibinfo {title} {Linear momentum in ferromagnets},\ }\href {https://doi.org/10.1088/0022-3719/20/7/003} {\bibfield  {journal} {\bibinfo  {journal} {J. Phys. C: Solid State Phys.}\ }\textbf {\bibinfo {volume} {20}},\ \bibinfo {pages} {L83} (\bibinfo {year} {1987})}\BibitemShut {NoStop}%
\bibitem [{\citenamefont {Berger}(1986)}]{berger_SMF}%
  \BibitemOpen
  \bibfield  {author} {\bibinfo {author} {\bibfnamefont {L.}~\bibnamefont {Berger}},\ }\bibfield  {title} {\bibinfo {title} {Possible existence of a {Josephson} effect in ferromagnets},\ }\href {https://doi.org/10.1103/PhysRevB.33.1572} {\bibfield  {journal} {\bibinfo  {journal} {Phys. Rev. B}\ }\textbf {\bibinfo {volume} {33}},\ \bibinfo {pages} {1572} (\bibinfo {year} {1986})}\BibitemShut {NoStop}%
\bibitem [{\citenamefont {Barnes}\ and\ \citenamefont {Maekawa}(2007)}]{barnes_SMF}%
  \BibitemOpen
  \bibfield  {author} {\bibinfo {author} {\bibfnamefont {S.~E.}\ \bibnamefont {Barnes}}\ and\ \bibinfo {author} {\bibfnamefont {S.}~\bibnamefont {Maekawa}},\ }\bibfield  {title} {\bibinfo {title} {Generalization of {Faraday}’s {Law} to {Include} {Nonconservative} {Spin} {Forces}},\ }\href {https://doi.org/10.1103/PhysRevLett.98.246601} {\bibfield  {journal} {\bibinfo  {journal} {Phys. Rev. Lett.}\ }\textbf {\bibinfo {volume} {98}},\ \bibinfo {pages} {246601} (\bibinfo {year} {2007})}\BibitemShut {NoStop}%
\bibitem [{\citenamefont {Yokouchi}\ \emph {et~al.}(2020)\citenamefont {Yokouchi}, \citenamefont {Kagawa}, \citenamefont {Hirschberger}, \citenamefont {Otani}, \citenamefont {Nagaosa},\ and\ \citenamefont {Tokura}}]{yokouchi_emergent_inductor}%
  \BibitemOpen
  \bibfield  {author} {\bibinfo {author} {\bibfnamefont {T.}~\bibnamefont {Yokouchi}}, \bibinfo {author} {\bibfnamefont {F.}~\bibnamefont {Kagawa}}, \bibinfo {author} {\bibfnamefont {M.}~\bibnamefont {Hirschberger}}, \bibinfo {author} {\bibfnamefont {Y.}~\bibnamefont {Otani}}, \bibinfo {author} {\bibfnamefont {N.}~\bibnamefont {Nagaosa}},\ and\ \bibinfo {author} {\bibfnamefont {Y.}~\bibnamefont {Tokura}},\ }\bibfield  {title} {\bibinfo {title} {Emergent electromagnetic induction in a helical-spin magnet},\ }\href {https://doi.org/10.1038/s41586-020-2775-x} {\bibfield  {journal} {\bibinfo  {journal} {Nature}\ }\textbf {\bibinfo {volume} {586}},\ \bibinfo {pages} {232} (\bibinfo {year} {2020})}\BibitemShut {NoStop}%
\bibitem [{\citenamefont {Wang}\ \emph {et~al.}(2023)\citenamefont {Wang}, \citenamefont {Zhang}, \citenamefont {Bheemarasetty}, \citenamefont {Ying},\ and\ \citenamefont {Xiao}}]{wang_singleSkNoise}%
  \BibitemOpen
  \bibfield  {author} {\bibinfo {author} {\bibfnamefont {K.}~\bibnamefont {Wang}}, \bibinfo {author} {\bibfnamefont {Y.}~\bibnamefont {Zhang}}, \bibinfo {author} {\bibfnamefont {V.}~\bibnamefont {Bheemarasetty}}, \bibinfo {author} {\bibfnamefont {S.-C.}\ \bibnamefont {Ying}},\ and\ \bibinfo {author} {\bibfnamefont {G.}~\bibnamefont {Xiao}},\ }\bibfield  {title} {\bibinfo {title} {Electronic noise of a single skyrmion},\ }\href {https://doi.org/10.1103/PhysRevB.108.094431} {\bibfield  {journal} {\bibinfo  {journal} {Phys. Rev. B}\ }\textbf {\bibinfo {volume} {108}},\ \bibinfo {pages} {094431} (\bibinfo {year} {2023})}\BibitemShut {NoStop}%
\bibitem [{\citenamefont {Marley}\ \emph {et~al.}(1995)\citenamefont {Marley}, \citenamefont {Higgins},\ and\ \citenamefont {Bhattacharya}}]{marley_fluxNoise}%
  \BibitemOpen
  \bibfield  {author} {\bibinfo {author} {\bibfnamefont {A.~C.}\ \bibnamefont {Marley}}, \bibinfo {author} {\bibfnamefont {M.~J.}\ \bibnamefont {Higgins}},\ and\ \bibinfo {author} {\bibfnamefont {S.}~\bibnamefont {Bhattacharya}},\ }\bibfield  {title} {\bibinfo {title} {Flux {Flow} {Noise} and {Dynamical} {Transitions} in a {Flux} {Line} {Lattice}},\ }\href {https://doi.org/10.1103/PhysRevLett.74.3029} {\bibfield  {journal} {\bibinfo  {journal} {Phys. Rev. Lett.}\ }\textbf {\bibinfo {volume} {74}},\ \bibinfo {pages} {3029} (\bibinfo {year} {1995})}\BibitemShut {NoStop}%
\bibitem [{\citenamefont {McMillan}(1976)}]{mcmillan_DCtheory}%
  \BibitemOpen
  \bibfield  {author} {\bibinfo {author} {\bibfnamefont {W.~L.}\ \bibnamefont {McMillan}},\ }\bibfield  {title} {\bibinfo {title} {Theory of discommensurations and the commensurate-incommensurate charge-density-wave phase transition},\ }\href {https://doi.org/10.1103/PhysRevB.14.1496} {\bibfield  {journal} {\bibinfo  {journal} {Phys. Rev. B}\ }\textbf {\bibinfo {volume} {14}},\ \bibinfo {pages} {1496} (\bibinfo {year} {1976})}\BibitemShut {NoStop}%
\bibitem [{\citenamefont {Rice}\ \emph {et~al.}(1981)\citenamefont {Rice}, \citenamefont {Whitehouse},\ and\ \citenamefont {Littlewood}}]{rice_impurityDC}%
  \BibitemOpen
  \bibfield  {author} {\bibinfo {author} {\bibfnamefont {T.~M.}\ \bibnamefont {Rice}}, \bibinfo {author} {\bibfnamefont {S.}~\bibnamefont {Whitehouse}},\ and\ \bibinfo {author} {\bibfnamefont {P.}~\bibnamefont {Littlewood}},\ }\bibfield  {title} {\bibinfo {title} {Impurity pinning of discommensurations in charge-density waves},\ }\href {https://doi.org/10.1103/PhysRevB.24.2751} {\bibfield  {journal} {\bibinfo  {journal} {Phys. Rev. B}\ }\textbf {\bibinfo {volume} {24}},\ \bibinfo {pages} {2751} (\bibinfo {year} {1981})}\BibitemShut {NoStop}%
\bibitem [{\citenamefont {Srolovitz}\ \emph {et~al.}(1987)\citenamefont {Srolovitz}, \citenamefont {Eykholt}, \citenamefont {Barnett},\ and\ \citenamefont {Hirth}}]{srolovitz_movingDC}%
  \BibitemOpen
  \bibfield  {author} {\bibinfo {author} {\bibfnamefont {D.~J.}\ \bibnamefont {Srolovitz}}, \bibinfo {author} {\bibfnamefont {R.}~\bibnamefont {Eykholt}}, \bibinfo {author} {\bibfnamefont {D.~M.}\ \bibnamefont {Barnett}},\ and\ \bibinfo {author} {\bibfnamefont {J.~P.}\ \bibnamefont {Hirth}},\ }\bibfield  {title} {\bibinfo {title} {Moving discommensurations interacting with diffusing impurities},\ }\href {https://doi.org/10.1103/PhysRevB.35.6107} {\bibfield  {journal} {\bibinfo  {journal} {Phys. Rev. B}\ }\textbf {\bibinfo {volume} {35}},\ \bibinfo {pages} {6107} (\bibinfo {year} {1987})}\BibitemShut {NoStop}%
\bibitem [{\citenamefont {Higgs}\ and\ \citenamefont {Gill}(1983)}]{higgs_hysteresis}%
  \BibitemOpen
  \bibfield  {author} {\bibinfo {author} {\bibfnamefont {A.~W.}\ \bibnamefont {Higgs}}\ and\ \bibinfo {author} {\bibfnamefont {J.~C.}\ \bibnamefont {Gill}},\ }\bibfield  {title} {\bibinfo {title} {Hysteresis in the electrical properties of orthorhombic tantalum trisulphide: {Evidence} for an incommensurate-commensurate charge-density wave transition?},\ }\href {https://doi.org/10.1016/0038-1098(83)90646-4} {\bibfield  {journal} {\bibinfo  {journal} {Solid State Commun.}\ }\textbf {\bibinfo {volume} {47}},\ \bibinfo {pages} {737} (\bibinfo {year} {1983})}\BibitemShut {NoStop}%
\bibitem [{\citenamefont {Parkin}\ \emph {et~al.}(2008)\citenamefont {Parkin}, \citenamefont {Hayashi},\ and\ \citenamefont {Thomas}}]{parkin_Racetrack}%
  \BibitemOpen
  \bibfield  {author} {\bibinfo {author} {\bibfnamefont {S.~S.~P.}\ \bibnamefont {Parkin}}, \bibinfo {author} {\bibfnamefont {M.}~\bibnamefont {Hayashi}},\ and\ \bibinfo {author} {\bibfnamefont {L.}~\bibnamefont {Thomas}},\ }\bibfield  {title} {\bibinfo {title} {Magnetic {Domain}-{Wall} {Racetrack} {Memory}},\ }\href {https://doi.org/10.1126/science.1145799} {\bibfield  {journal} {\bibinfo  {journal} {Science}\ }\textbf {\bibinfo {volume} {320}},\ \bibinfo {pages} {190} (\bibinfo {year} {2008})}\BibitemShut {NoStop}%
\bibitem [{\citenamefont {Tomasello}\ \emph {et~al.}(2014)\citenamefont {Tomasello}, \citenamefont {Martinez}, \citenamefont {Zivieri}, \citenamefont {Torres}, \citenamefont {Carpentieri},\ and\ \citenamefont {Finocchio}}]{tomasello_SkLRacetrack}%
  \BibitemOpen
  \bibfield  {author} {\bibinfo {author} {\bibfnamefont {R.}~\bibnamefont {Tomasello}}, \bibinfo {author} {\bibfnamefont {E.}~\bibnamefont {Martinez}}, \bibinfo {author} {\bibfnamefont {R.}~\bibnamefont {Zivieri}}, \bibinfo {author} {\bibfnamefont {L.}~\bibnamefont {Torres}}, \bibinfo {author} {\bibfnamefont {M.}~\bibnamefont {Carpentieri}},\ and\ \bibinfo {author} {\bibfnamefont {G.}~\bibnamefont {Finocchio}},\ }\bibfield  {title} {\bibinfo {title} {A strategy for the design of skyrmion racetrack memories},\ }\href {https://doi.org/10.1038/srep06784} {\bibfield  {journal} {\bibinfo  {journal} {Sci. Rep.}\ }\textbf {\bibinfo {volume} {4}},\ \bibinfo {pages} {6784} (\bibinfo {year} {2014})}\BibitemShut {NoStop}%
\end{thebibliography}%
\end{document}


\title{Supplemental Material for:\\Current-induced sliding motion in a helimagnet MnAu$_2$}

\author{Yuta~Kimoto}
\email{yuta.kimoto.t7@dc.tohoku.ac.jp}
\author{Hidetoshi~Masuda}
\affiliation{Institute for Materials Research, Tohoku University, Sendai 980-8577, Japan}
\author{Takeshi~Seki}
\affiliation{Institute for Materials Research, Tohoku University, Sendai 980-8577, Japan}
\affiliation{Center for Science and Innovation in Spintronics (CSIS), Core Research Cluster, Tohoku University, Sendai 980-8577, Japan}
\author{Yoichi~Nii}
\affiliation{Institute for Materials Research, Tohoku University, Sendai 980-8577, Japan}
\author{Jun-ichiro~Ohe}
\affiliation{Department of Physics, Toho University, Funabashi 274-8510, Japan}
\author{Yusuke~Nambu}
\affiliation{Institute for Materials Research, Tohoku University, Sendai 980-8577, Japan}
\author{Yoshinori~Onose}
\affiliation{Institute for Materials Research, Tohoku University, Sendai 980-8577, Japan}

\maketitle
\tableofcontents
\newpage

\section{Analysis of ferromagnetic transition field} 
\begin{figure}[ht]
  \includegraphics[width=17.2cm]{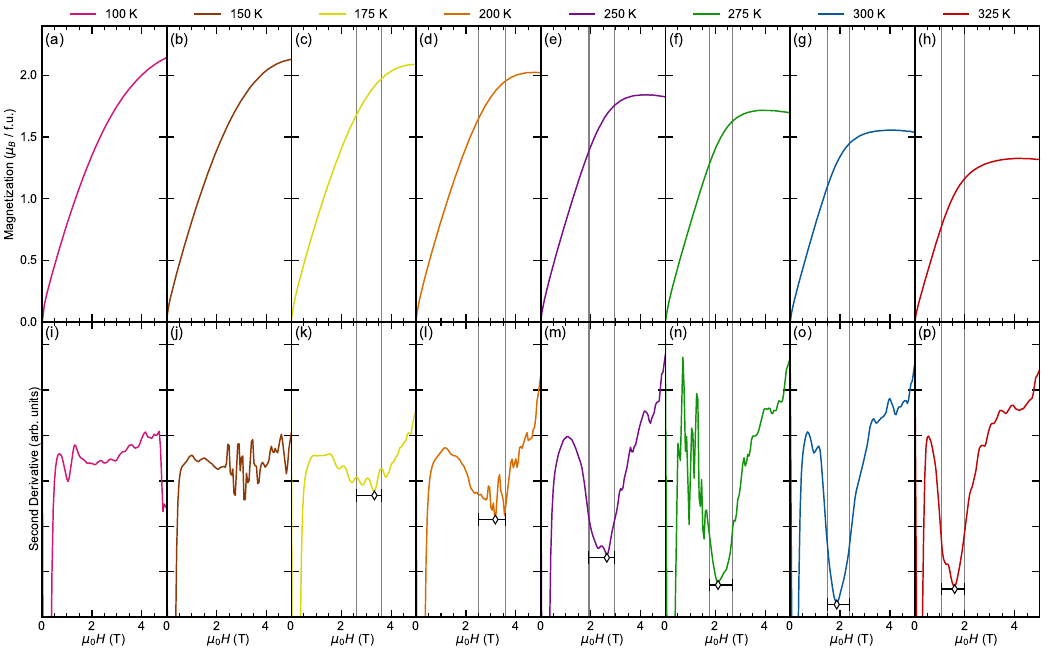}
  \caption{(a)-(h) Magnetization curves at various temperatures in the $\mathrm{Mn}\mathrm{Au}_2$ thin film. The measurement was done using a Magnetic Property Measurement System (Quantum Design). The sample has a rectangular shape. The longer side along the [001] direction was 8.0 mm, and the shorter side along the [1-10] direction was 4 mm. The thickness was 100 nm. The diamagnetic component of $\mathrm{Sc}\mathrm{Mg}\mathrm{Al}\mathrm{O}_4$ substrate is subtracted. (i)-(p) The second derivative of the magnetization curves. The negative broad peak is caused by the conical-to-ferromagnetic transition. In the thin film, the transition is slightly broad, possibly due to the epitaxial strain. The transition field is estimated above 175 K as the peak and the width is estimated as the width of two-thirds maximum, which is plotted as the error bars in these figures as well as in Fig. 3(b) in the main text and \autoref{FigS7}. Below 150 K, we cannot identify distinguishable peak structures from the second derivative of the magnetization curves.}
  \label{FigS1}
\end{figure}
\newpage

\section{$\Delta\rho_{xx}$ and low-frequency noise in the chirality-uncontrolled state}
\begin{figure}[ht]
  \includegraphics[height=9.0cm]{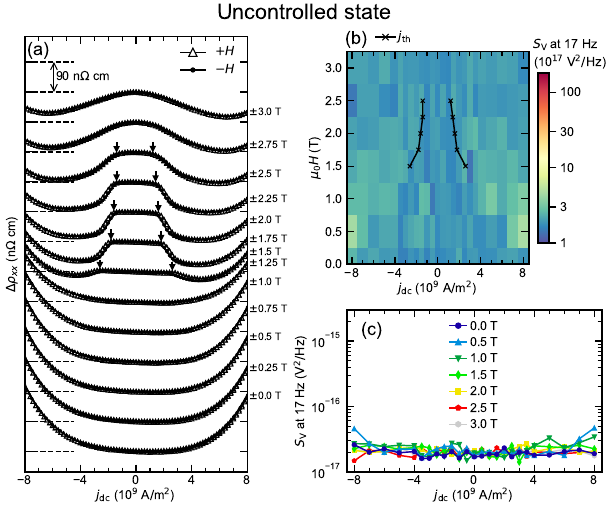}
  \caption{Differential resistivity and voltage noise for the chirality-uncontrolled state at 250 K. (a) $\Delta\rho_{xx}$ at various magnetic fields. The black arrows denote threshold currents $j_\mathrm{th}$. (b) The magnetic field dependence of $j_\mathrm{th}$. Contour mapping of voltage noise at 17 Hz as functions of $j_\mathrm{dc}$ and magnetic field is superimposed. (c) The power spectral density of voltage noise at 17 Hz as a function of $j_\mathrm{dc}$ under various magnetic fields.}
  \label{FigS2}
\end{figure}

Figure \ref{FigS2} shows $\Delta\rho_{xx}$ and voltage noise at various magnetic fields at 250 K for the chirality uncontrolled state. While $\Delta\rho_{xx}$ shows kinks at the threshold currents above 1.5 T, we could not discern any anomaly in the lower field region. Voltage noise was not observed in the whole magnetic field region. For the chirality-controlled state, voltage noise was enhanced in the low field region, where the sliding is suppressed for the chirality-uncontrolled state. That is one of the reasons why the voltage noise was not observed in the chirality-uncontrolled state.

\newpage

\section{Effect of shape anisotropy and/or epitaxial strain on $\Delta\rho_{xx}$}
\begin{figure}[ht]
  \includegraphics[height=9.0cm]{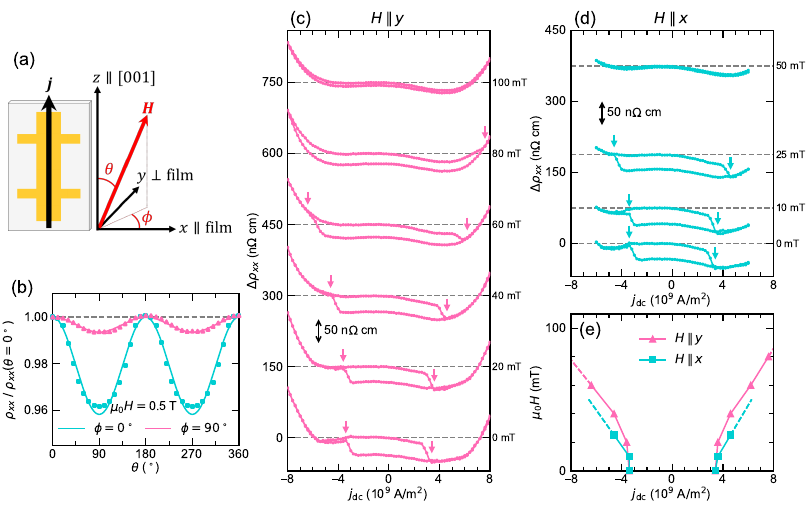}
  \caption{(a) Schematic image of the film and the magnetic field directions. (b) Normalized resistivity as a function of $\theta$ for $\phi=0^\circ$ and $\phi=90^\circ$ at 250 K and 0.5 T. Here, $\theta$, $\phi$ are spherical coordinates of the magnetic field shown in (a). The magnetic field vector is expressed as $(H\sin\theta\cos\phi,H\sin\theta\sin\phi,H\cos\theta)$. (c, d) $\Delta\rho_{xx}$ for $+$chiral state at 250 K at various magnetic fields for (c) $H\parallel y$ and (d) $H\parallel x$. The coefficient $\alpha$ used in the derivation of $\Delta\rho_{xx}$ (Eq. (1) in the main text) is $1.25\times10^{-15} \mathrm{A}^{-2}\mathrm{m}^4$ for both the cases. The arrows represent threshold currents $j_\mathrm{th}$. (e) The magnetic field dependence of the threshold currents $j_\mathrm{th}$ at 250 K for the two magnetic field directions: $H\parallel y$ (triangles), and $H\parallel x$ (squares).}
  \label{FigS3}
\end{figure}

In \autoref{FigS3}, we show $\rho_{xx}$ and $\Delta\rho_{xx}$ in magnetic fields parallel to the (001) plane to study the effect of shape anisotropy and epitaxial strain. As shown in \autoref{FigS3}(a), the magnetic field direction is expressed by spherical coordinates $\theta$, $\phi$ or orthogonal coordinates $x$, $y$, $z$. Figure \ref{FigS3}(b) shows $\theta$ dependence of resistivity at $\phi=0^\circ$ and $\phi=90^\circ$. While $\rho_{xx}$ shows sinusoidal $\theta$ dependence with the period of 180 degrees for both $\phi=0^\circ$ and $\phi=90^\circ$, $\rho_{xx}$ at $\theta=90^\circ$ and $\phi=90^\circ$ is larger than that at $\theta=90^\circ$ and $\phi=0^\circ$, indicating finite epitaxial strain and/or shape anisotropy. For the measurement of $\Delta\rho_{xx}$, we first controlled the chirality to $+$chiral state by using the electric current and a magnetic field along the [001] direction. Then, we rotated the sample by 90$^\circ$ and measured $\Delta\rho_{xx}$ in magnetic fields perpendicular to the [001] direction. The results are shown in Figs. \ref{FigS3}(c, d). The threshold currents are much larger than those in magnetic fields along the [001] direction shown in the main text. The magnetic field dependence of the threshold current for $H\parallel x$ and $H\parallel y$ is plotted in \autoref{FigS3}(e). The threshold current for $H\parallel y$ is larger than that for $H\parallel x$, which is caused by the epitaxial strain and/or shape anisotropy.

\newpage

\section{Nonreciprocal electronic transport for two metastable states in the low-field hysteresis}
\begin{figure}[ht]
  \includegraphics[width=9.0cm]{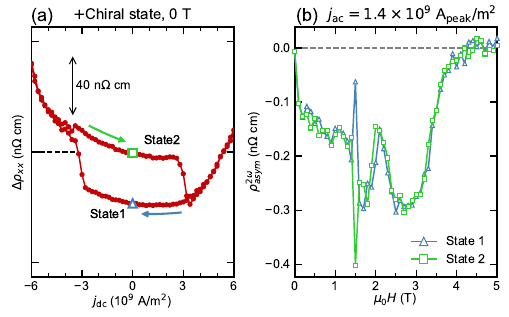}
  \caption{(a) $\Delta\rho_{xx}$ at 250 K and 0T reproduced from Fig. 3(a) in the main text. State 1 and State 2 represent two metastable states, in which we measured nonreciprocal electronic transport.  (b) Nonreciprocal electronic transport for State 1 and State 2.}
  \label{FigS4}
\end{figure}

As discussed in the main text, $\Delta\rho_{xx}$ shows hysteresis in the low magnetic field region, which seems to be caused by some metastable states. To elucidate whether the hysteresis is caused by the chirality degree of freedom or not, we measured the nonreciprocal electronic transport for two metastable states. At first, we applied positive and negative electric currents and turned them off to realize the metastable states [\autoref{FigS4}(a)]. Then, we measured the nonreciprocal electronic transport [\autoref{FigS4}(b)]. The signs are both positive and therefore the chirality is not relevant to the hysteresis phenomenon.

\newpage

\section{Comparison of Real and Imaginary parts of $\rho_{xx}$\\at various magnetic fields}
\begin{figure}[ht]
  \includegraphics[width=17.2cm]{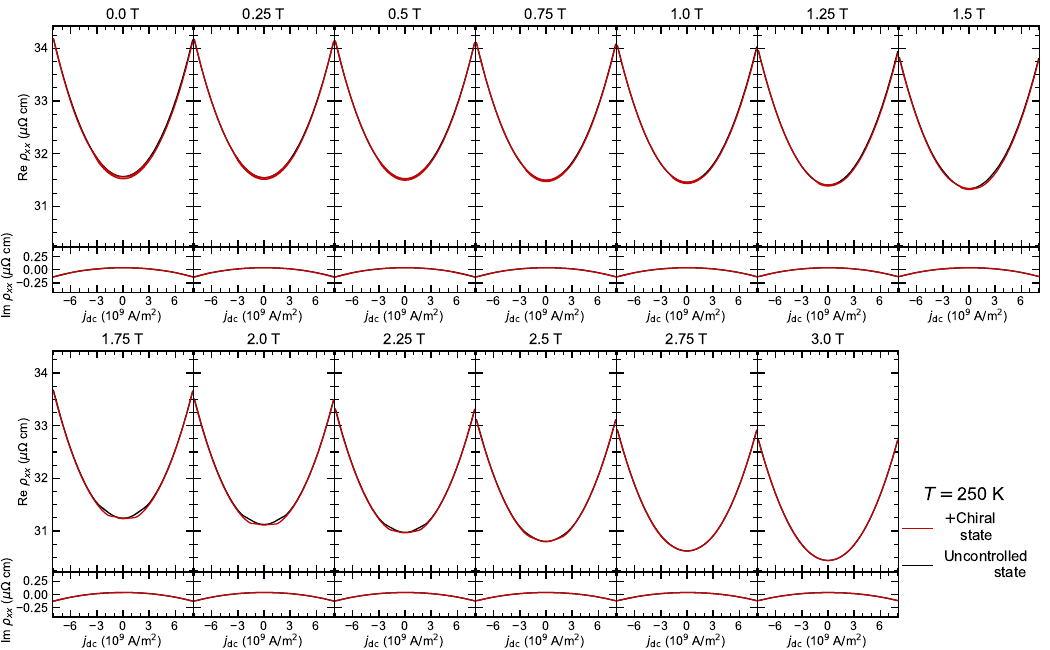}
  \caption{The real and imaginary parts of $\rho_{xx}$ as a function of $j_\mathrm{dc}$ at 250 K. The magnitude of the imaginary part is smaller than 0.5\% of the real part for all the magnetic field data.}
  \label{FigS5}
\end{figure}
\newpage

\section{Temperature dependence of $\Delta\rho_{xx}$ and $j_\mathrm{th}$ }
\begin{figure}[h]
  \includegraphics[width=17.2cm]{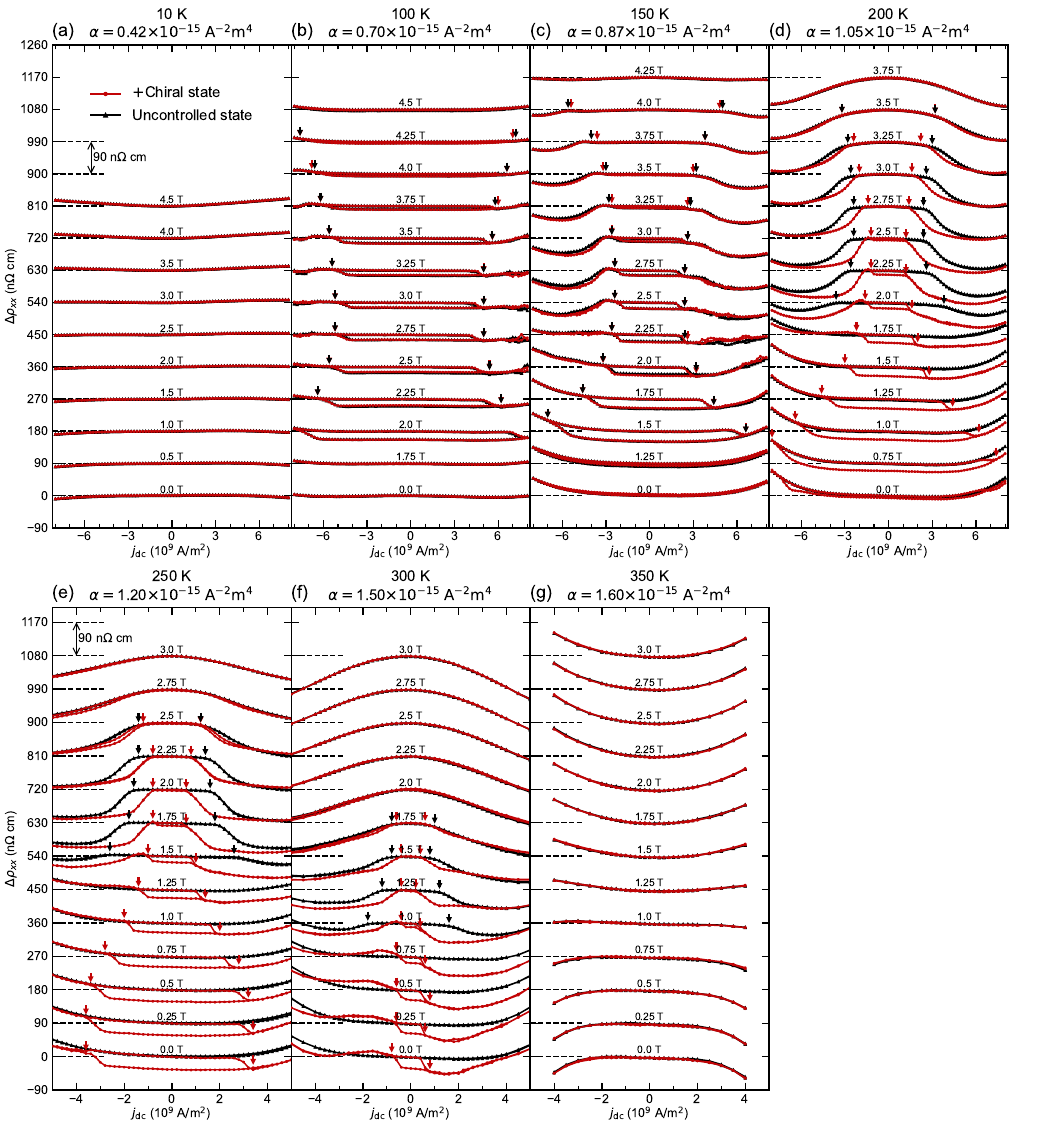}
  \caption{$\Delta\rho_{xx}$ as a function of $j_\mathrm{dc}$ at various magnetic fields and temperatures. The red and black lines represent the data for $+$chiral and uncontrolled states, respectively. The arrows indicated in each panel represent threshold currents $j_\mathrm{th}$. The coefficient $\alpha$ used in the derivation of $\Delta\rho_{xx}$ (Eq. (1) in the main text) is determined for each temperature so as to minimize the current dependence of $\Delta\rho_{xx}$ in the high current region for all the field data.}
  \label{FigS6}
\end{figure}
\newpage

\begin{figure}[h]
  \includegraphics[width=17.2cm]{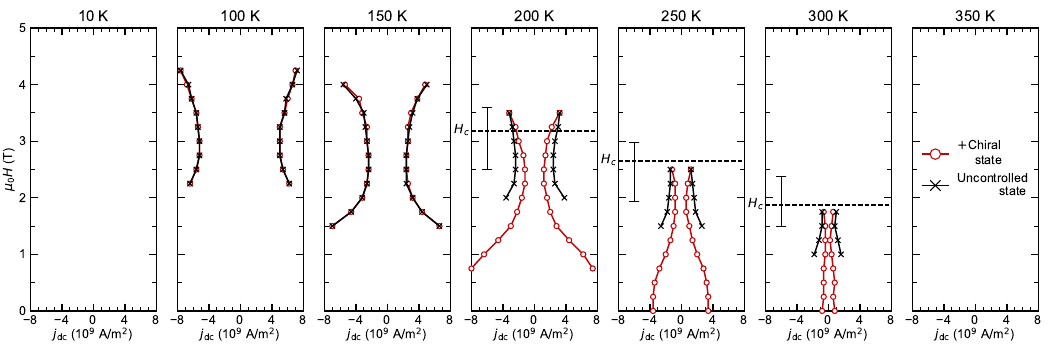}
  \caption{Temperature and magnetic field dependence of threshold currents $j_\mathrm{th}$. The dashed lines shown for 200 K, 250 K, and 300 K denote the conical-to-ferromagnetic phase transition fields. The transitions are broad in this thin film sample presumably due to the epitaxial strain and the width of the transition field is shown as an error bar (for the details, see \autoref{FigS1}).}
  \label{FigS7}
\end{figure}

\section{Optical microscope image of the sample}
\begin{figure}[ht]
  \includegraphics[width=8.6cm]{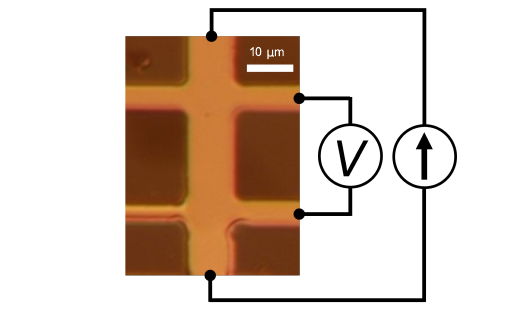}
  \caption{Optical microscope image of the $\mathrm{Mn}\mathrm{Au}_2$ thin film sample for the measurement of data shown in the main text. The sample was shaped into a Hall bar with the use of lithography technique and no obvious scar was discerned.}
  \label{FigS8}
\end{figure}
\newpage

\section{Sample dependence}
To show the sample variation, we have measured two additional samples (\#2, \#3). We have measured differential resistivity $\rho_{xx}$ and voltage noise $S_\mathrm{V}$ for \#2 while only $\rho_{xx}$ was measured for \#3. For \#2, the abrupt change of $\Delta\rho_{xx}$ was not discerned in the low magnetic field region while the broadband noise emerged in the high current region even at lower magnetic fields. For \#3, we have observed abrupt change of $\rho_{xx}$ at the threshold current and asymmetric $\Delta\rho_{xx}$ in the low magnetic field region.

\subsection{Sample \#2}
\begin{figure}[htb]
  \includegraphics[height=9.0cm]{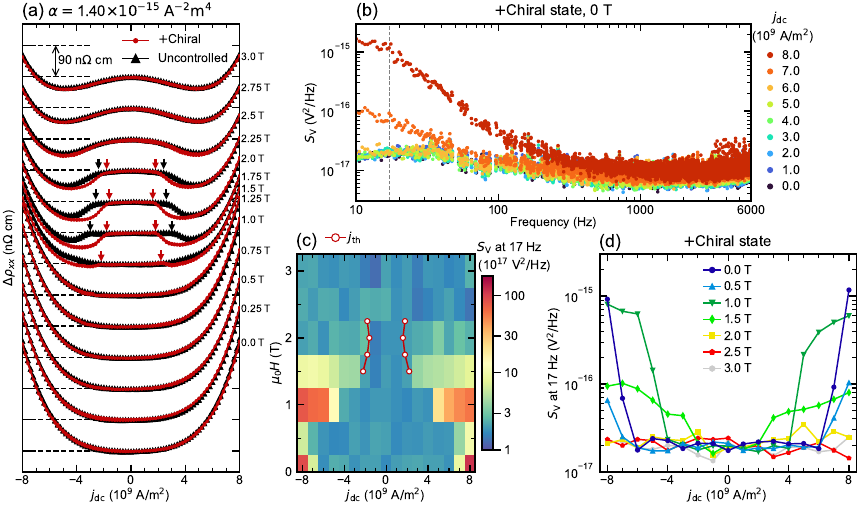}
  \caption{Differential resistivity and voltage noise for sample \#2 at 250 K. (a) $\Delta\rho_{xx}$ for $+$chiral and uncontrolled states under various magnetic fields. The arrows represent threshold currents $j_\mathrm{th}$. The coefficient $\alpha$ used in the derivation of $\Delta\rho_{xx}$ (Eq. (1) in the main text) is determined so as to minimize current dependence of $\Delta\rho_{xx}$ in the high current region for all the field data. (b) Voltage noise spectra under various $j_\mathrm{dc}$ for $+$chiral state at 0 T. (c) The magnetic field dependence of $j_\mathrm{th}$ for $+$chiral state. Contour mapping of voltage noise at 17 Hz is superimposed. (d) The voltage noise at 17 Hz as a function of $j_\mathrm{dc}$ under various magnetic fields at 250 K.}
  \label{FigS9}
\end{figure}
\newpage

\subsection{Sample \#3}
\begin{figure}[htb]
  \includegraphics[width=7.0cm]{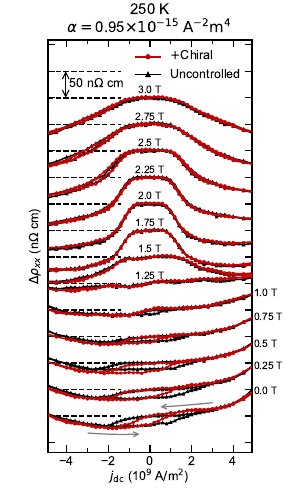}
  \caption{Differential resistivity for sample \#3 at 250 K. $\Delta\rho_{xx}$ for $+$chiral and uncontrolled states at various magnetic fields are plotted as circles and triangles, respectively. The coefficient $\alpha$ used in the derivation of $\Delta\rho_{xx}$ (Eq. (1) in the main text) is determined so as to minimize current dependence of $\Delta\rho_{xx}$ in the high current region for all the field data.}
  \label{FigS10} 
\end{figure}
\newpage

\section{Analysis of noise spectrum}
\begin{figure}[ht]
  \includegraphics[width=17.2cm]{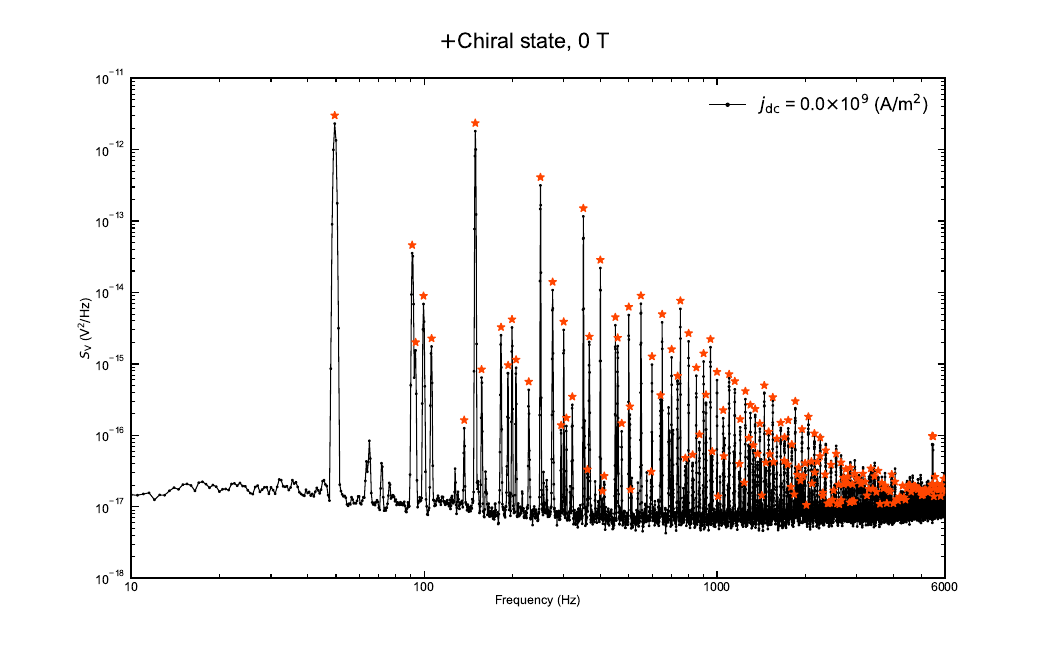}
  \caption{The raw voltage spectrum for $+$chiral state at 0 T. The orange star symbols indicate extrinsic background noises observed even when DC bias current is absent. These sharp structures are removed for all noise spectra displayed in our work.}
  \label{FigS12}
\end{figure}

In our measurement environment, there are extrinsic voltage noises. Figure \ref{FigS12} shows the voltage noise spectra without any bias current. The spiky peaks originated from the extrinsic noise. In the analysis of voltage noise spectra, we have removed these spiky peaks.

\newpage

\section{Noise spectra at various magnetic fields}
\begin{figure}[ht]
  \includegraphics[width=17.2cm]{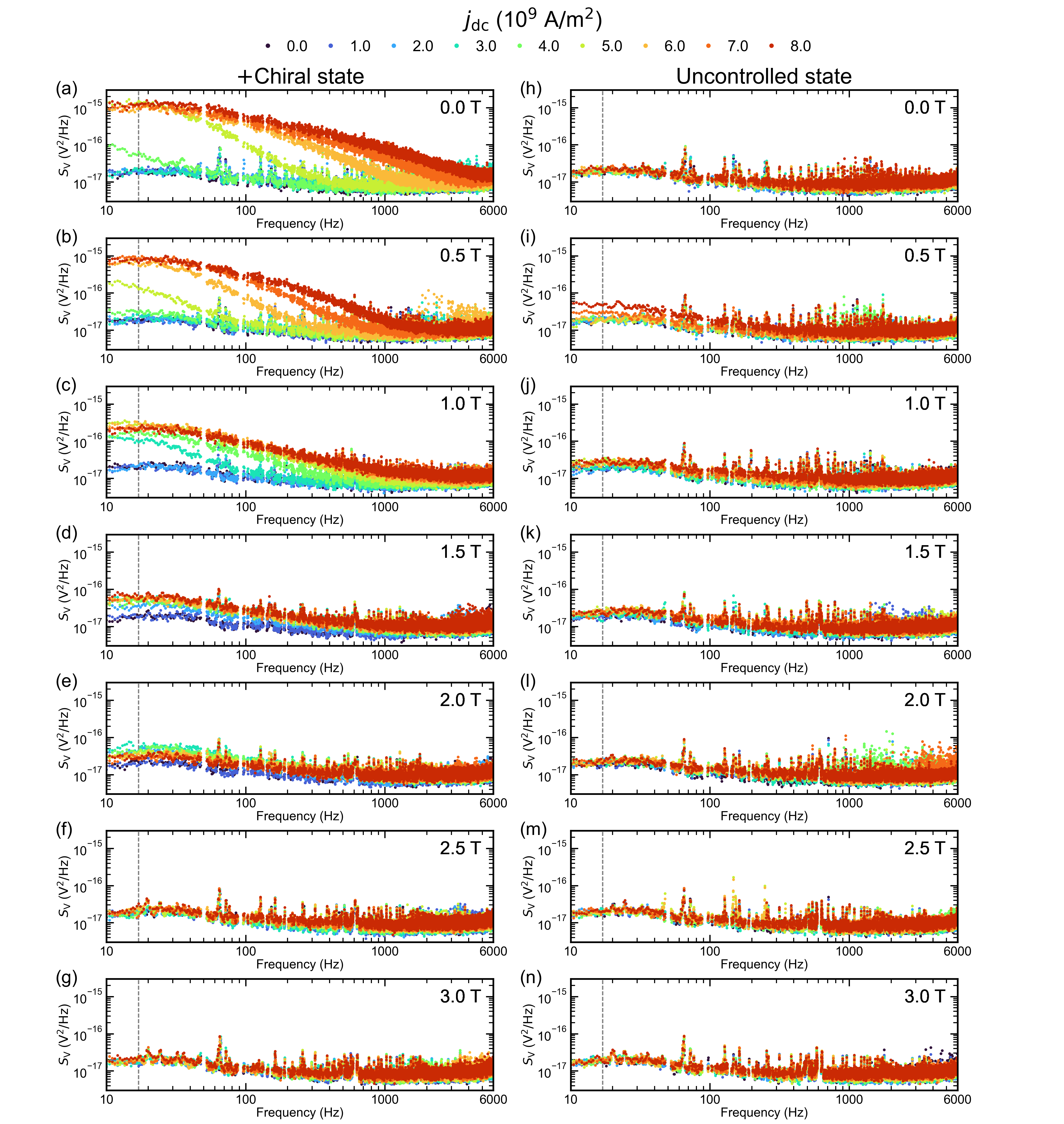}
  \caption{(a)-(n) The voltage noise spectra at 250 K under various $j_\mathrm{dc}$ and magnetic fields for (a)-(g) $+$chiral state and (h)-(n) uncontrolled state.}
  \label{FigS13}
\end{figure}